\documentclass[preprint,doublespacing,12pt,authoryear]{elsarticle}

\usepackage[utf8]{inputenc} 
\usepackage[T1]{fontenc} 
\usepackage[english]{babel}
\usepackage{multicol}
\usepackage{graphicx}
\usepackage{amsmath}
\usepackage{amssymb}
\usepackage[hidelinks]{hyperref}
\addto\extrasenglish{%

}
\usepackage{booktabs}
\usepackage{xcolor}
\usepackage{tabularx}
\usepackage{lineno}
\usepackage{csquotes}
\usepackage[section]{placeins}
\usepackage{float}
\usepackage{threeparttable}
\usepackage[margin=2.5cm]{geometry}
\usepackage{setspace}
\usepackage{caption} 
\begin{document}

\begin{frontmatter}

\journal{Ecological Complexity}



\title{Predatory dynamics in susceptible and resistant \textit{Eriopis connexa} populations}


\author[inst1]{Anna Mara Ferreira Maciel}
\author[inst2, inst3]{Gabriel Rodrigues Palma}
\author[inst4]{Lucas dos Anjos}
\author[inst1]{Lucas Santos Canuto}
\author[inst1]{Wesley Augusto Conde Godoy}
\author[inst2]{Rafael de Andrade Moral}

\affiliation[inst1]{organization={Department of Entomology and Acarology, University of São Paulo, ESALQ},
            addressline={Av. Padua Dias, 11 - Agronomia}, 
            city={Piracicaba},
            postcode={13418-900}, 
            state={São Paulo},
            country={Brazil}}
\affiliation[inst2]{organization={Department of Mathematics and Statistics},
            addressline={Maynooth University}, 
            city={Maynooth, Co. Kildare},
            postcode={W23 F2H6}, 
            country={Ireland}}
\affiliation[inst3]{organization={Hamilton Institute},
            addressline={Maynooth University}, 
            city={Maynooth, Co. Kildare},
            postcode={W23 AH3Y}, 
            country={Ireland}}
\affiliation[inst4]{organization={East China Normal University},
            addressline={No. 500 Dongchuan Road, Minhang District}, 
            city={Shanghai},
            postcode={200241}, 
            country={China}}
\begin{abstract}
The ladybird \textit{Eriopis connexa} (Germar, 1824), a voracious aphid predator, faces challenges from insecticide applications, compromising biological control. As a result, there has been an increase in the number of studies analysing the resistance and susceptibility of ladybirds. Some studies have found that resistant populations exhibit distinct predation and foraging behaviour compared to susceptible ones. This study models the population dynamics of resistant and susceptible \textit{E. connexa} preying on \textit{Aphis gossypii} Glover, 1877 and \textit{Myzus persicae} (Sulzer, 1776). We constructed a logistic model with density dependence and type-II functional response to analyse predation dynamics, incorporating bifurcation analysis on predation parameters (attack rate and handling time) and the mortality rate of susceptible ladybirds. We simulated scenarios with/without insecticide application and with/without aphid resistance. To simulate the effects of insecticide applications, the parameters related to aphids' intrinsic growth rate ($r_1$ and $r_2$) change to reflect the responses of susceptible and resistant populations. The same approach is used concerning the mortality rate of ladybirds ($d_2$ and $d_3$). Our results demonstrate that mortality, attack rate, and handling time are critical in shaping predator-prey interactions. Temporal simulations revealed fluctuating abundances, highlighting the fragility of these interactions under insecticide stress. Therefore, this study contributes to understanding the ecological implications of insecticides, which disrupt natural predation dynamics, and shows how variations in behavioural rates can impact prey control. This research demonstrated the importance of integrated strategies that balance insecticide applications with preserving natural enemies and promoting sustainable agricultural practices. 

\end{abstract}

\begin{graphicalabstract}
\centering
\includegraphics[width = 0.5\textwidth]{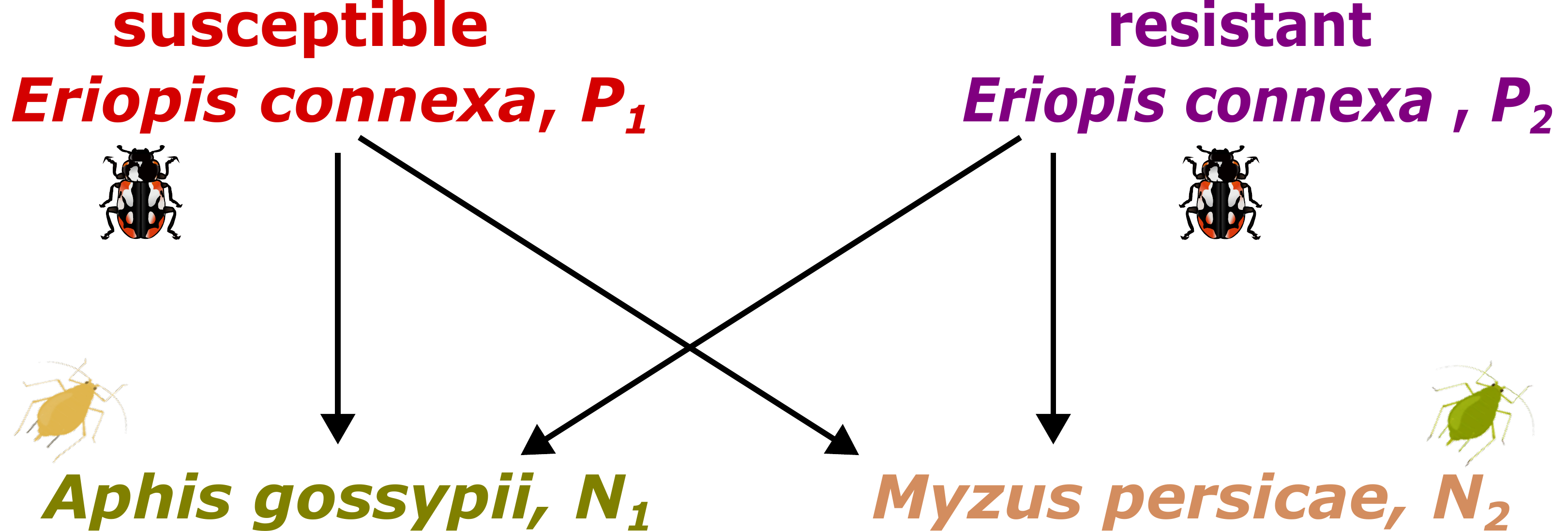}
\end{graphicalabstract}

\sloppy
\begin{highlights}
\item Coexistence of susceptible and resistant predators acts as ecological insurance under insecticide stress.
\item Mathematical simulations show that insecticides disrupt aphid-ladybird dynamics and highlight the importance of integrated pest management.
\end{highlights}
\fussy


\end{frontmatter}


\section{Introduction}
Controlling pests through combined synthetic insecticides and natural enemies is a fundamental strategy in Integrated Pest Management (IPM) programs. Although insecticides remain a key tool for pest suppression, their indiscriminate use can affect ecological interactions, causing consequences such as pest resistance evolution, effects on non-target insects, and pest resurgence \citep{agathokleous2023low, kho2024control, lira2024assessment}. Insecticides can fail to control aphids due to resistance, while they can reduce natural enemies’ populations, causing pest outbreaks. This phenomenon, often called pest resurgence, shows the complexity of managing pest populations. Furthermore, the difficulty of increasing the use of natural enemies is the constant need for insecticide applications to control pests, compromising the survival of these agents. Mathematical models predict that long-term equilibrium densities of pests will increase with insecticide applications when the predators also suffer from increased mortality \citep{janssen2021pesticides, trumper1998modelling}. 

Resistance is not exclusive to pests, and natural enemies can also develop resistant phenotypes to certain insecticides \citep{bielza2016insecticide, cheng2022transcriptome,rodrigues2013response}. This can help pest management combine insecticides that are selective to natural enemies and negatively affect pests. The presence of susceptible and resistant populations of natural enemies can influence predator-prey dynamics. An effect may include the delay of resistance development by pests targeted with insecticides since surviving individuals that are likely to be resistant could be consumed by natural enemies resistant to insecticides \citep{lira2019predation}. Although resistant populations can survive in treated insecticide environments, they also can show sublethal effects such as changes in foraging behaviour and predatory efficiency \citep{cloyd2012indirect, desneux2007sublethal}. While susceptible populations can be more effective predators in insecticide-free environments, when insecticides are applied, they become vulnerable, which increases their mortality.

Although there is growing recognition of these interactions, limited research exists on the long-term population dynamics of predator-prey systems and the predation behaviour of resistant natural enemies when insecticides are applied, particularly considering the coexistence with susceptible natural enemies, their increased mortality, and the resulting ecological effects. The use of mathematical models allows simulations and analysis of complex dynamics. When integrating data about foraging behaviour, predation rates, and functional response of predators in different scenarios, models can predict the impacts of different management strategies for predator-prey populations. These models are particularly valuable to test hypotheses and scenarios that could be logistically difficult or unfeasible to realize experimentally \citep{ferreira2014ecological}.

Mathematical modelling is a tool with applications in biological control, providing valuable insights into interactions between insects and agricultural practices, with the resulting ideas and frameworks having broad ecological application \citep{alexandridis2021models, moral2023modelling, mcevoy2018theoretical}. Furthermore, previous studies have used mathematical models to explore scenarios involving insecticide applications in predator-prey systems, and other studies have incorporated resistance evolution in pests or natural enemies \citep{janssen2021pesticides, trumper1998modelling, tabashnik1986evolution, liu2021modelling, jana2013mathematical, tang2005integrated} but in different contexts from the one we address here. In this study, we integrate resistance dynamics, considering fitness costs and changes in the predatory efficiency of resistant predators into a specific ecological context. Our approach evaluates how varying resistance scenarios and insecticide applications influence the interaction between the aphids \textit{Aphis gossypii} Glover, 1877 and \textit{Myzus persicae} (Sulzer, 1776) and their predator, the ladybird \textit{Eriopis connexa} (Germar, 1824).

Aphids are among the most challenging pests to control in various cropping systems because of their rapid reproductive cycles, selection for resistant phenotypes, and potential to vector plant pathogens \citep{emden2017aphids, blackman2000aphids}. Here, we chose to study \textit{M. persicae} and \textit{A. gossypii} because they are two polyphagous species that infest a wide variety of crops \citep{blackman2000aphids}. We selected the ladybird \textit{E. connexa} as their natural enemy. On a global scale, it serves as an important biological control agent against aphids and mites, distinguished by its polyphagy, high voracity, and natural presence in several economically important crops \citep{sarmento2007functional, nascimento2021performance, fidelis2018coccinellidae,silva2013biological,grez2005ladybird}. Recent studies have identified an insecticide-resistant phenotype in this species, with documented occurrence of a fitness cost and changes in predatory potential for these individuals \citep{ferreira2013, rodrigues2016ontogenic, wang2020feeding, lira2019predation, tavares2010selective}. 

Natural enemies can be exposed to insecticides while preying on pests, causing resistant and susceptible predators to have different predatory efficiencies. Their presence may influence pest population dynamics and control strategies. Therefore, we aim to determine whether the presence of resistant \textit{E. connexa} can mitigate \textit{A. gossypii} and \textit{M. persicae} outbreaks in insecticide-treated environments and explore the implications for sustainable IPM strategies. Using a mathematical modelling approach within this context, we simulate different conditions to assess how the evolution of resistance in both aphids and their predators influences pest control outcomes. We hypothesise that (1) insecticide-resistant populations of \textit{E. connexa} will persist in scenarios with insecticide application, contributing to aphid population control as susceptible conspecifics decline; (2) scenarios involving aphid resistance to insecticides will make pest control more difficult and may lead to pest resurgence over time; (3) parameter such as attack rate and handling time are important in predator-prey stability, helping maintain predator populations in the system and contributing to pest control; and (4) insecticide application can disrupt the predator-prey equilibria, potentially leading to aphid outbreaks due to predator suppression and the differential recovery rates between aphids and their predators.

\section{Methods}

\subsection{Model and Simulations}

\subsubsection{Assumptions}

\begin{figure*} [ht]
    \centering
    \includegraphics[width = 0.5\textwidth]{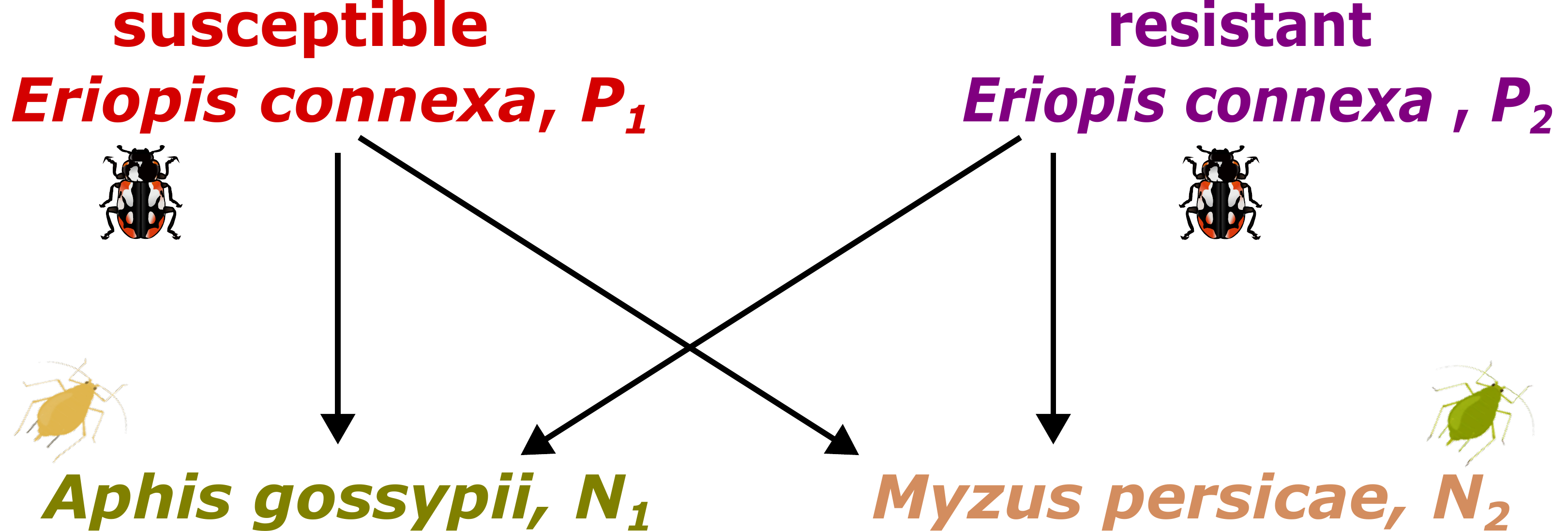}
    \caption{Diagram illustrating the biological system studied, which includes the resistant and susceptible predators and aphid species.}
    \label{fig:scheme}
\end{figure*}

We developed a mathematical model consisting of two plant pests and a natural enemy with two subpopulations, susceptible and resistant to insecticides. We assumed the density-dependent logistic model, incorporating a multi-species type II functional response to model prey consumption. The model is non-structured in terms of life stages, and we also assumed that \textit{E. connexa} does not have any preference between the two aphid species.

The first assumption of the model is that resistant populations of ladybirds bear a cost in fitness and predation efficiency \citep{ferreira2013,rodrigues2016ontogenic, lira2019predation}. Additionally, resistant \textit{E. connexa} individuals exhibit differences in attack rates and handling times compared to susceptible ones \citep{lira2019predation}. We separated ladybird populations into two equations due to parameter differences between susceptible and resistant individuals. The presence of insecticide resistance can alter multiple parameters, including mortality and predation efficiency, expressed as attack rate and handling time. These additional parameters require a more detailed model to differentiate the behaviour of susceptible and resistant individuals. 

The functional response of \textit{E. connexa} in our model is assumed to be a FR type II. This assumption is supported by the findings of  \cite{sarmento2007functional}. Although the functional response (FR) is a relevant relationship in studying the dynamics of predator-prey models, the growth rate is also an essential demographic parameter, as it encompasses birth and mortality rates, indicating whether a population is expanding or declining. It becomes particularly important under the influence of insecticides, commonly used in agriculture to manage pest populations. Insecticides can influence the intrinsic growth rate by affecting survival and reproduction, thus shaping population dynamics over time. Different species tend to exhibit different growth rates when exposed to insecticides, which impacts population density and selective pressure on resistant genotypes \citep{wang2020feeding}. It is crucial to ensure that resistant aphid species genotypes are selected under selection pressure and, in tropical regions, exhibit parthenogenetic reproduction. Therefore, when resistance is selected, subsequent generations will also be resistant \citep{margaritopoulos2021long}. Unlike ladybirds, for which separate equations are used to model susceptible and resistant populations, a single equation was used for each aphid species since only the intrinsic growth rate is affected by insecticides. Within this equation, a function adjusts the growth rate based on insecticide exposure, effectively capturing its impact on the aphid population.

In scenarios with the application of insecticides selecting for resistant aphid individuals, the intrinsic growth rate increases with each application. However, the magnitude of this increase reduces progressively with subsequent applications, reflecting a decreasing impact over time. In scenarios without the selection of resistant individuals, the intrinsic growth rate decreases, and the effectiveness of the insecticide intensifies with repeated applications.

The effects of insecticides were modelled through two mechanisms: a logistic function applied to the intrinsic growth rates of aphids ($r_1$ and $r_2$) (\autoref{eq:growth_rate}) and an exponential function used to influence the mortality rates of ladybirds ($d_2$ and $d_3$) (\autoref{eq:mortality}). The host crop was excluded from the model, and the host's influence on the parameters was considered constant throughout the simulation. Incorporating the above assumptions, a system of ordinary differential equations to model the population dynamics of this system was written as:

\begin{equation}\label{eq:aphis_gossypi}
\begin{split}
\underbrace{\frac{dN_1}{dt}}_{\text{\textit{Aphis gossypii}}} 
= & \underbrace{r_1(t) N_1 \left(\frac{K_1 - N_1}{K_1} \right)}_{\text{logistic growth of pest } N_1} 
 - \underbrace{\left(\frac{a^{(s)}_1 N_1}{1 + a^{(s)}_1 N_1 T_{h_1}^{(s)} + a^{(s)}_2 N_2 T_{h_2}^{(s)}}\right)}_{\text{multispecies FR of } P_1 \text{ upon } N_1} P_1 \\
& - \underbrace{\left(\frac{a^{(r)}_1 N_1}{1 + a^{(r)}_1 N_1 T_{h_1}^{(r)} + a^{(r)}_2 N_2 T_{h_2}^{(r)}} \right)}_{\text{multispecies FR of agent } P_2 \text{ upon pest } N_1} P_2 
\end{split}
\end{equation}

\begin{equation}\label{eq:myzus_persicae}
\begin{split}
\underbrace{\frac{dN_2}{dt}}_{\text{\textit{Myzus persicae}}} 
= & \underbrace{r_2(t) N_2 \left(\frac{K_2 - N_2}{K_2} \right)}_{\text{logistic growth of pest } N_2} - \underbrace{\left(\frac{a^{(s)}_2 N_2}{1 + a^{(s)}_2 N_2 T_{h_2}^{(s)} + a^{(s)}_1 N_1 T_{h_1}^{(s)}}\right)}_{\text{multispecies FR of } P_1 \text{ upon } N_2} P_1 \\
& - \underbrace{\left(\frac{a^{(r)}_2 N_2}{1 + a^{(r)}_2 N_2 T_{h_2}^{(r)} + a^{(r)}_1 N_1 T_{h_1}^{(r)}} \right)}_{\text{multispecies FR of } P_2 \text{ upon } N_2} P_2 
\end{split}
\end{equation}

\begin{equation}\label{eq:susceptible_ladybird}
\begin{split}
\underbrace{\frac{dP_1}{dt}}_{\text{susceptible \textit{Eriopis connexa}}} 
= & \underbrace{-d_2(t) P_1^2}_{\text{quadratic mortality rate of } P_1} \\
& + \underbrace{\left(\frac{g^{(s)}_1 a^{(s)}_1 N_1 + g^{(s)}_2 a^{(s)}_2 N_2}{1 + a^{(s)}_1 N_1 T_{h_1}^{(s)} + a^{(s)}_2 N_2 T_{h_2}^{(s)}} \right)}_{\text{multispecies numerical response of } P_1 \text{ from consumption of } N_1 \text{ and } N_2} P_1 
\end{split}
\end{equation}

\begin{equation}\label{eq:resistant_ladybird}
\begin{split}
\underbrace{\frac{dP_2}{dt}}_{\text{resistant \textit{Eriopis connexa}}} 
= & \underbrace{-d_3(t) P_2^2}_{\text{quadratic mortality rate of agent } P_2} \\
& + \underbrace{\left(\frac{g^{(r)}_1 a^{(r)}_1 N_1 + g^{(r)}_2 a^{(r)}_2 N_2}{1 + a^{(r)}_1 N_1 T_{h_1}^{(r)} + a^{(r)}_2 N_2 T_{h_2}^{(r)}} \right)}_{\text{multispecies numerical response of } P_2 \text{ from consumption of } N_1 \text{ and } N_2} P_2 
\end{split}
\end{equation}

$N_1$ and $N_2$ are the densities of the pests (prey) (Eqs.~\ref{eq:aphis_gossypi} and \ref{eq:myzus_persicae}), while $P_1$ and $P_2$ represent the densities of the control agents (predators) (Eqs.~\ref{eq:susceptible_ladybird} and \ref{eq:resistant_ladybird}). The variables' definitions and the model's parameters are shown in Table~\ref{tab:my-table1}. 

\begin{table*}[ht]
\centering
\caption{Definition of variables and parameters used in the mathematical model, representing resistant and susceptible predators and aphid species}
\label{tab:my-table1}
\begin{threeparttable}
\begin{tabular}{@{}cc@{}}
\toprule
\multicolumn{2}{c}{\textbf{Meaning}}                                                       \\ \midrule
\textbf{Variables}  &                                                                      \\
$N_1$                  & Density of \textit{Aphis gossypii}                                            \\
$N_2$                  & Density of \textit{Myzus persicae}                                            \\
$P_1$                  & Density of susceptible \textit{Eriopis connexa}                                \\
$P_2$                  & Density of resistant \textit{Eriopis connexa}                                  \\
\textbf{Parameters} &                                                                      \\
$r_1$                  & Intrinsic growth rate of $N_1$                              \\
$r_2$                  & Intrinsic growth rate of $N_2$                             \\
$K_1$                  & Carrying capacity of $N_1$                                  \\
$K_2$                  & Carrying capacity of $N_2$                                 \\
$a^{(s)}_1$                 & Attack coefficient of $P_1$ on pest $N_1$                                  \\
$a^{(s)}_2$                 & Attack coefficient of $P_1$ on pest $N_2$                                  \\
$a^{(r)}_1$                 & Attack coefficient of $P_2$ on pest $N_1$                                  \\
$a^{(r)}_2$                 & Attack coefficient of $P_2$ on pest $N_2$                                  \\
$T_{h_1}^{(s)}$                & Handling time of agent $P_1$ on pest $N_1$                             \\
$T_{h_2}^{(s)}$                & Handling time of agent $P_1$ on pest $N_2$                             \\
$T_{h_1}^{(r)}$                & Handling time of agent $P_2$ on pest $N_1$                             \\
$T_{h_2}^{(r)}$                & Handling time of agent $P_2$ on pest $N_2$                             \\
$d_2$                  & Coefficient of density-dependent per capita mortality rate of agent $P_1$             \\
$d_3$                  & Coefficient of density-dependent per capita mortality rate of agent $P_2$            \\
$g^{(s)}_1$                 & Conversion coefficient from pest $N_1$ to agent $P_1$                      \\
$g^{(s)}_2$                 & Conversion coefficient from pest $N_2$ to agent $P_1$                      \\
$g^{(r)}_1$                 & Conversion coefficient from pest $N_1$ to agent $P_2$                      \\
$g^{(r)}_2$                 & Conversion coefficient from pest $N_2$ to agent $P_2$                      \\ \bottomrule
\end{tabular}
   \begin{tablenotes}
            \footnotesize
            \item [] $r_1$, $K_1$ = parameters related to \textit{Aphis gossypii}; $r_2$, $K_2$ = parameters related to \textit{Myzus persicae}; $a^{(s)}_1$, $a^{(s)}_2$, $T_{h_1}^{(s)}$, $T_{h_2}^{(s)}$, $d_2$, $g^{(s)}_1$, $g^{(s)}_2$ = parameters related to susceptible \textit{Eriopis connexa}; $a^{(r)}_1$, $a^{(r)}_2$, $T_{h_1}^{(r)}$, $T_{h_2}^{(r)}$, $d_3$, $g^{(r)}_1$, $g^{(r)}_2$ = parameters related to resistant \textit{Eriopis connexa}.
        \end{tablenotes}
    \end{threeparttable}
\end{table*}

The first equation describes the growth rate of the \textit{A. gossypii} population ($N_1$) over time, while the second represents \textit{M. persicae} ($N_2$). The initial terms in both equations account for population growth based on the intrinsic rate ($r_1$ and $r_2$) and carrying capacity ($K_1$ and $K_2$). The remaining terms describe the effect of the type II functional response, which depends on the attack rate parameters ($a_1$ and $a_2$) and handling time ($T_{h1}$ and $T_{h2}$). Here, we used the multispecies functional response model without assuming interspecific exploitative competition between the pests.

The function designed to vary the intrinsic growth rate proposes that after the insecticide application, the values of ($r$) are modified, shifting from the initial values corresponding to the susceptible populations to those of the resistant populations. When there is no insecticide application, the values of $r_1$ and $r_2$ remain the same as those defined in the initial conditions. However, when an application occurs, the mathematical function is triggered, resulting in a decrease in the values of $r_1$ and $r_2$ close to the application time, followed by a slow recovery until the new values of $r_1$ and $r_2$ are reached. In our model, we assume that in scenarios where aphid resistance is selected, the final $r$-values are higher than the initial ones. This assumption is based on findings from \cite{wang2020feeding,ma2019fitness}, which indicates that resistant populations exhibit higher intrinsic growth rates than susceptible ones. This increase in $r$-values is incorporated into the model to reflect the adaptive advantage observed in resistant populations. 

Conversely, in scenarios where resistance is not selected, we assume that $r$-values decrease following insecticide applications. This assumption is supported by studies such as \cite{liu2022sublethal, qiu2024intergenerational}, which report negative effects of insecticide exposure on the growth rates of susceptible populations. To capture this dynamic, the model includes a functional decrease in the $r$-values after insecticide application, simulating the costs associated with exposure. 

In the model, the function is defined as shown in \autoref{eq:growth_rate}:

\begin{equation} \label{eq:growth_rate}
r_i(t) =
\begin{cases} 
r_{i0}, & t \leq t_{\text{app}},\ i \in \{1,2\}, \\
\begin{aligned}[t]
  &-\alpha + e^{-\beta \cdot t} \\
  &\quad + S\left(t, r_{\mathit{new},i}, t_{\text{app}} + \theta, \frac{1}{\gamma}\right),
\end{aligned} & t > t_{\text{app}},\ i \in \{1,2\}.
\end{cases}
\end{equation}

where \( r_{i0} \) represents the initial value of $r$ for population  \( i \), \( t_{\text{app}} \) is the insecticide application time, \( r_{\text{new},i} \) corresponds to the new $r$ value for population  \( i \). The parameter $\alpha$ represents the growth level, $\beta$ is the decline speed, $\theta$ defines the (\({delay}\)) between application and the onset of changes, $\gamma$ determines the growth speed of $r$, and $S(t)$ is a logistic function. In other words, they determine the transition dynamics.

The third and fourth equations describe the rate of change of the susceptible ($P_1$) and resistant ($P_2$) \textit{E. connexa} subpopulations. In both aspects of the equations, converting consumed prey ($N_1$ and $N_2$) into new predators within the same time interval involves the structure of the type II functional response and the conversion rate. However, in the first part of the equations, the effect of the conspecific predator's mortality rate is included, which has a function to vary the parameter's value according to insecticide application.

The exponential function models a process where the value does not change when there is no insecticide application. However, after this point, the value begins to grow exponentially until it reaches a new value while simultaneously undergoing an exponential decline that starts with a certain delay. In this way, the function only increases the mortality rate when insecticide applications occur and then returns to values close to the natural mortality rate. The increase in value for susceptible ladybirds is higher than for resistant ones. Thus, the function for the conspecific predator’s mortality rate is given by \autoref{eq:mortality}:

\begin{equation} \label{eq:mortality}
d_i(t) =
\begin{cases} 
d_{i0}, & t \leq t_{\text{app}},\ i \in \{1,2\}, \\[8pt]
\begin{aligned}[t]
d_{i0} &+ (d_{i0} + (d_{\text{new},i} - d_{i0}) \cdot e^{-\gamma \cdot (t-t_{\text{app}})} \\
&\quad - \alpha \cdot e^{-\beta \cdot (t-(t_{\text{app}}+\theta))},
\end{aligned} & t > t_{\text{app}},\ i \in \{1,2\}
\end{cases}
\end{equation}

where:
$d_{i0}$ is the initial mortality rate for population \( i \) (before insecticide application),
\( d_{\text{new},i} \) is the mortality rate after insecticide application,
\( t_{\text{app}} \) is the insecticide application time. After application, the parameters $\alpha$ (\({growth\_level}\)), $\beta$ (\({decline\_speed}\)), $\theta$ ((\({delay}\)), and $\gamma$ (\({growth\_speed}\)) control the transition dynamics.

This function ensures that before insecticide application, the mortality rate remains constant, and after application, it dynamically adjusts to reflect the expected changes in mortality rate.

\subsubsection{Initial conditions and parameter values} 
\label{subsec:variables}

The initial densities in the standard simulations were $30$ individuals for each pest species ($N_1$ and $N_2$), $10$ susceptible ladybirds ($P_1$), and 5 resistant ladybirds ($P_2$). Most parameter values for simulations were taken from a literature survey focusing on the species of aphids as the pest insect and ladybirds, especially \textit{E. connexa} as the predator insect. 

Estimated values for intrinsic growth rate ($r_1$ and $r_2$) were available for the studies from \textit{A. gossypii} and \textit{M. persicae}. The first focused on cotton being the host plant \citep{kerns2000sublethal}, and for \textit{M. persicae} a range of host plants was researched because no study was found in cotton. Therefore, we selected an intermediate value within this range that closely matched and best represented the data found based on the study by \cite{wang2018laboratory}.

The model parameters for the ladybirds were primarily derived from the study of a pest–predator system involving \textit{E. connexa} and mites by \cite{matos2020eriopis}. While the current model considers aphids as prey, we also reviewed the literature on other Coccinellidae species preying on aphids to ensure that the parameter values closely align with those observed for mite-based systems \citep{lee2004306, jalali2017effects, bayoumy2018foraging, moradi2020foraging}. This cross-referencing helped validate the applicability of the selected values, maintaining ecological relevance across prey types. 

To estimate the parameters for resistant \textit{E. connexa}, we referenced the study carried out by \cite{lira2019predation}. This study reported that resistant ladybirds exhibited approximately a $46\%$ reduction in attack rate and an $8\%$ decrease in handling time compared to susceptible individuals. Accordingly, our 
model incorporates these same percentage reductions to reflect the behaviour of resistant \textit{E. connexa}.

In this study, we assume that conspecific predation among predators primarily reduces their growth rates. A direct method to model the mortality caused by conspecific killing in predators is through quadratic terms such as $d_2P_1^2$ e $d_3P_2^2$, as suggested by \cite{lucasintraguild}. The quadratic formulation effectively captures the non-linear increase in mortality as predator density rises. However, it is noteworthy that $d_2$ and $d_3$ represent only the mortality due to conspecific predation and do not account for insecticide-induced mortality. The mortality caused by insecticide application is expressed by an exponential function, which is activated only during insecticide application periods. 

Therefore, as the literature predominantly provides mortality rate values for predators without explicitly accounting for conspecific predation, we used bifurcation analysis to investigate the sensitivity of this parameter. To reflect density dependence in the absence of empirical estimates, we assigned an arbitrary value to the mortality rate associated with conspecific predation. Various ecological factors, including cannibalism, can influence the density-dependent mortality rate of predators, making this assumption necessary for model exploration. 

It is well-documented that cannibalism rates among predators increase when aphids are absent or their density is low \citep{grez2012biotic, cannibalism2017}. Experimental evidence of cannibalism in  \textit{E. connexa} is documented in \cite{cannibalism2017, rocca2020larval}. To account for this, our model incorporates a baseline mortality rate that includes cannibalism-driven losses, and when insecticides are applied, we increase this value to reflect the compound effects. The collapse of an aphid population due to insecticide application is expected to intensify competition for food in ladybirds, leading to a rise in conspecific predation. This dynamic effectively increases the quadratic mortality rate, emphasizing the interplay between food scarcity and intraspecific interactions in such scenarios. 

To modulate the induced mortality of ladybirds, we adjusted parameters following a similar approach to that described by  \cite{survival_2013}, which reported survival rates of $3\%$ for susceptible ladybirds and $84\%$ for resistant ones after insecticide exposure.  Therefore, these survival rates were used as proxies to approximate the effects of insecticide-induced mortality, as they primarily reflect external factors.

\subsection{Bifurcation} \label{subsec:bifurcation}
A numerical bifurcation analysis was conducted by varying selected model parameters while keeping the others constant, with a focus on the density-dependent mortality of the susceptible ladybird and the predation parameters of both subpopulations. A detailed description of the methodology and the full results are provided in the Appendix.

\subsection{Population Dynamics}\label{subsec:population}

In addition to bifurcation diagrams, we also simulated a range of scenarios to better understand the population dynamics generated by this system. Different from bifurcation, here we simulated fixing the quadratic mortality parameter of the susceptible ladybird ($d_2$). 

We conducted a simulation with $d_2 = 0.5$, characterized by sustained oscillations in the population dynamics. With sustained oscillations, this scenario offers a dynamic system that is sufficiently stable but still responds clearly to changes in parameters or initial conditions, allowing us to assess how patterns of population persistence and extinction emerge from interactions in the system. Thus, it allows for a more explicit and detailed analysis of changes in population dynamics when we introduce applications or disturbances to the system. By focusing on this oscillatory behaviour, we highlight how different factors directly influence the patterns of interaction in a species and reinforce the importance of dynamic coupling to understand the system's responses as a whole.

The simulations were conducted over $220$ time units, which represent days, in software R $4.4.1$ \citep{softwareR}. We selected the value 220 because, in general, the cotton cycle lasts a maximum of $220$ days. In scenarios involving insecticide applications, these applications began after $30$ days of simulation, in which no immigration or emigration of individuals was assumed. The insecticide was considered to be applied in the field as a pulse application for one day once every twenty days. The simulated scenarios were: (a) without insecticide application; (b) with insecticide application and without aphid resistance, and (c) with insecticide application and with aphid resistance. All simulations were run in R \citep{softwareR} with the aid of package \texttt{simecol} \citep{simecol}, using \textit{lsoda} as a solver. 

\subsection{Analysis of Population Dynamics}

Since estimated attack rates vary in the literature, we chose to investigate if those different values can disrupt biological control, especially between the two pest species. We also analysed parameters related to the predator's attack rate, such as the handling time. To analyse the effects of each parameter on the stability of the dynamics, we systematically varied the values of one parameter while keeping others fixed, as described in the subsection \ref{subsec:population}. Again, we chose the scenario without insecticide application. The analysed parameter spaces were defined as follows: $a^{(s)}_1 \in \{0, 0.01, \ldots , 1\}$,  
$a^{(s)}_2 \in \{0, 0.01, \ldots , 1\}$,  
$a^{(r)}_1 \in \{0, 0.01, \ldots , 1\}$,  
$a^{(r)}_2 \in \{0, 0.01, \ldots , 1\}$,  
$T_{h_1}^{(s)} \in \{0, 0.01, \ldots , 1\}$,  
$T_{h_2}^{(s)} \in \{0, 0.01, \ldots , 1\}$,  
$T_{h_1}^{(r)} \in \{0, 0.01, \ldots , 1\}$   
and  
$T_{h_2}^{(r)} \in \{0, 0.01, \ldots , 1\}$.. This approach allows us to assess how variations in each parameter influence the population sizes in the long term. By analysing the final population sizes across different parameter values, we can determine whether the system reaches equilibrium, exhibits fluctuations, or shows trends toward extinction or dominance of certain species. During each analysis, only the parameter of interest was varied, while all other parameters and variables remained fixed with the values specified in the subsection \ref{subsec:variables}.

\section{Results}
\subsection{Bifurcation}
The bifurcation analysis identified patterns indicating regions of stability and instability that are sensitive to the parameters analysed. See the appendix for full details.

\subsection{Population Dynamics}

\begin{figure*} [ht]
    \centering
    \includegraphics[width = 0.8\textwidth]{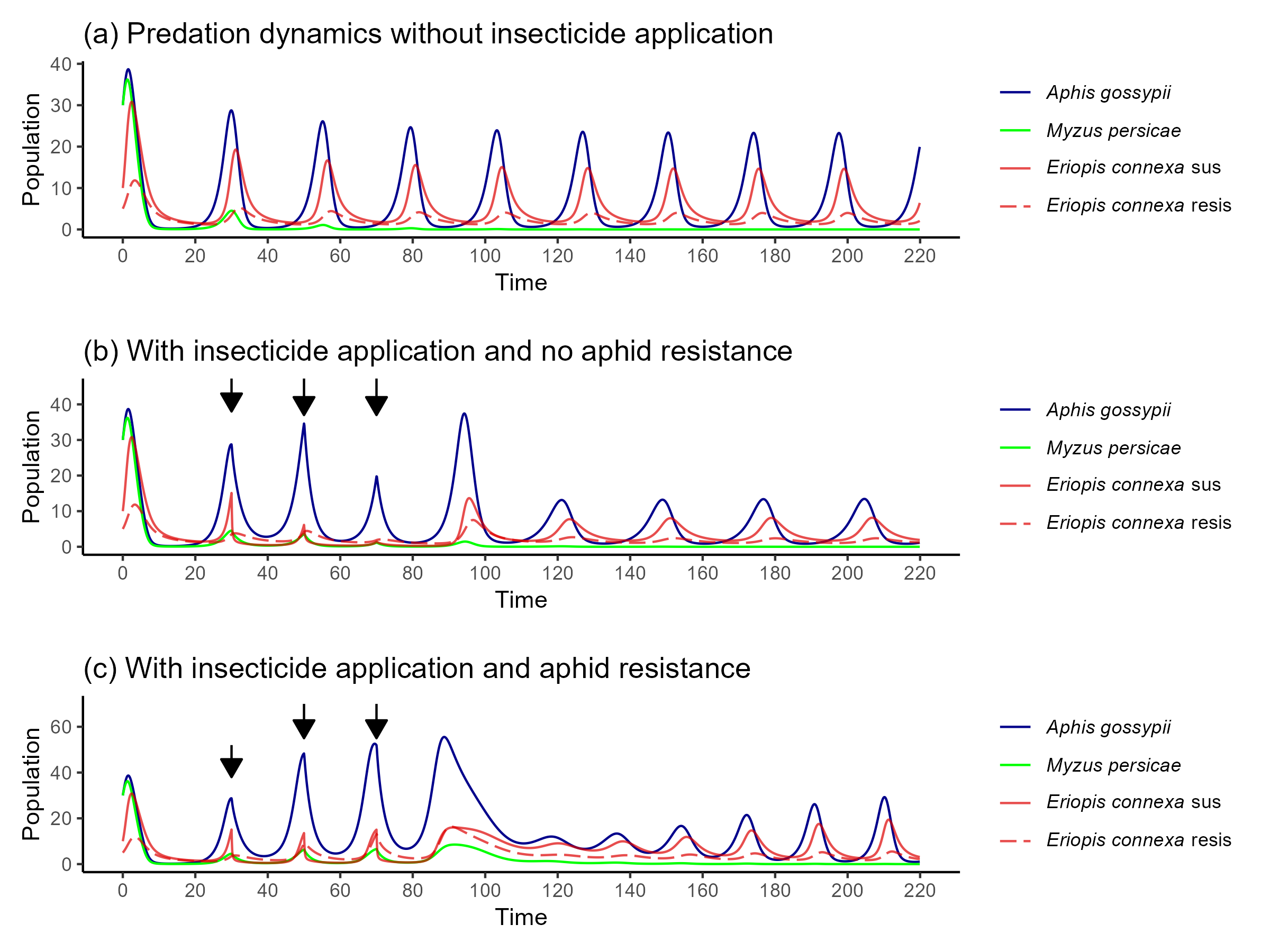} 
    \caption{Simulations of scenarios involving pest-predator dynamics: (a) without insecticide application, (b) with insecticide application and no aphid resistance, and (c) with insecticide application and aphid resistance. Initial conditions and parameter values: $N_1 = N_2 = 30$; $P_1 = 10$; $P_2 = 5$; $K_1 = K_2 = 200$; $a^{(s)}_1 = 0.20$; $a^{(s)}_2 = 0.21$; $T_{h_1}^{(s)} = 0.68$; $T_{h_2}^{(s)} = 0.73$; $a^{(r)}_1 = 0.10$; $a^{(r)}_2 = 0.11$; $T_{h_1}^{(r)} = 0.62$; $T_{h_2}^{(r)} = 0.67$; $d_2 = d_3 = 0.05$; $g^{(s)}_1 = g^{(s)}_2 = 0.2$; $g^{(r)}_1 = g^{(r)}_2 = 0.17$. In scenarios with insecticide application, the values of $r_1$, $r_2$, $d_2$, and $d_3$ are modified. The arrows indicate insecticide applications at 30, 50, and 70 days.}
    \label{fig:image10}
\end{figure*}

\autoref{fig:image10}a presents a scenario containing susceptible and resistant populations of ladybirds that was simulated without the application of insecticides to assess ladybird predation. It is noticeable that, because absence of the insecticide's applications in the system, the resistant ladybird population is not selected and tends toward extinction due to adaptive costs (\autoref{fig:image10}a). 

In \autoref{fig:image10}b, the application of insecticides was combined with the presence of ladybirds, without selection of aphid resistance, and in \autoref{fig:image10}c with resistance. In the second scenario (\autoref{fig:image10}b), a sharp decline in the populations of \textit{A. gossypii} (blue line) and \textit{M. persicae} (green line) occurs shortly after application (Time $30$, $50$, and $70$). Aphid mortality is evident, with populations reaching levels close to zero in the initial moments after the intervention. 
However, after this initial reduction, pest density subsequently increases over time, resulting in higher population levels. Additionally, there is a decline in predator populations of \textit{E. connexa} (solid and dashed red lines) due to prey shortages and insecticide application. However, predator populations gradually recover over time, particularly the resistant ladybirds, which increase through the applications and persist even when the susceptible ones decline. This leads to more effective pest control in later cycles, reestablishing the balance of predator-prey dynamics (\autoref{fig:image10}b). 

In the scenario in which aphids are resistant to the insecticide (\autoref{fig:image10}c), it can be seen that the applications result in a less pronounced initial reduction in the populations of \textit{A. gossypii} and \textit{M. persicae}. This reduction is followed by rapid population growth of aphids, which reaches densities higher than those observed in the scenario without insecticide application (\autoref{fig:image10}a). Regarding the populations of \textit{E. connexa}, the resistant ladybirds (dashed red line) show significant growth, maintaining population levels close to those of the susceptible ladybirds (solid red line). This response is because of the lower mortality of resistant ladybirds compared to the susceptible ones, favouring their permanence in the system even after multiple insecticide applications (\autoref{fig:image10}c).

After the peak of aphid populations, ladybird populations (susceptible and resistant) can temporarily reduce aphid densities, demonstrating the potential for controlling this natural predator even in an environment with high pest pressure. However, when insecticide applications cease, resistant ladybird populations begin to decline gradually, while the \textit{A. gossypii} population that remained in the environment began to grow sharply again. This increase results in new population peaks of \textit{A. gossypii}, indicating a scenario of resurgence of pests in the long term, mainly because of the combination of aphid resistance and the reduction of predatory ladybird populations (\autoref{fig:image10}c).

\FloatBarrier

\subsection{Analysis of Population Dynamics}

\begin{figure*} [ht]
    \centering
    \includegraphics[width = 0.8\textwidth]{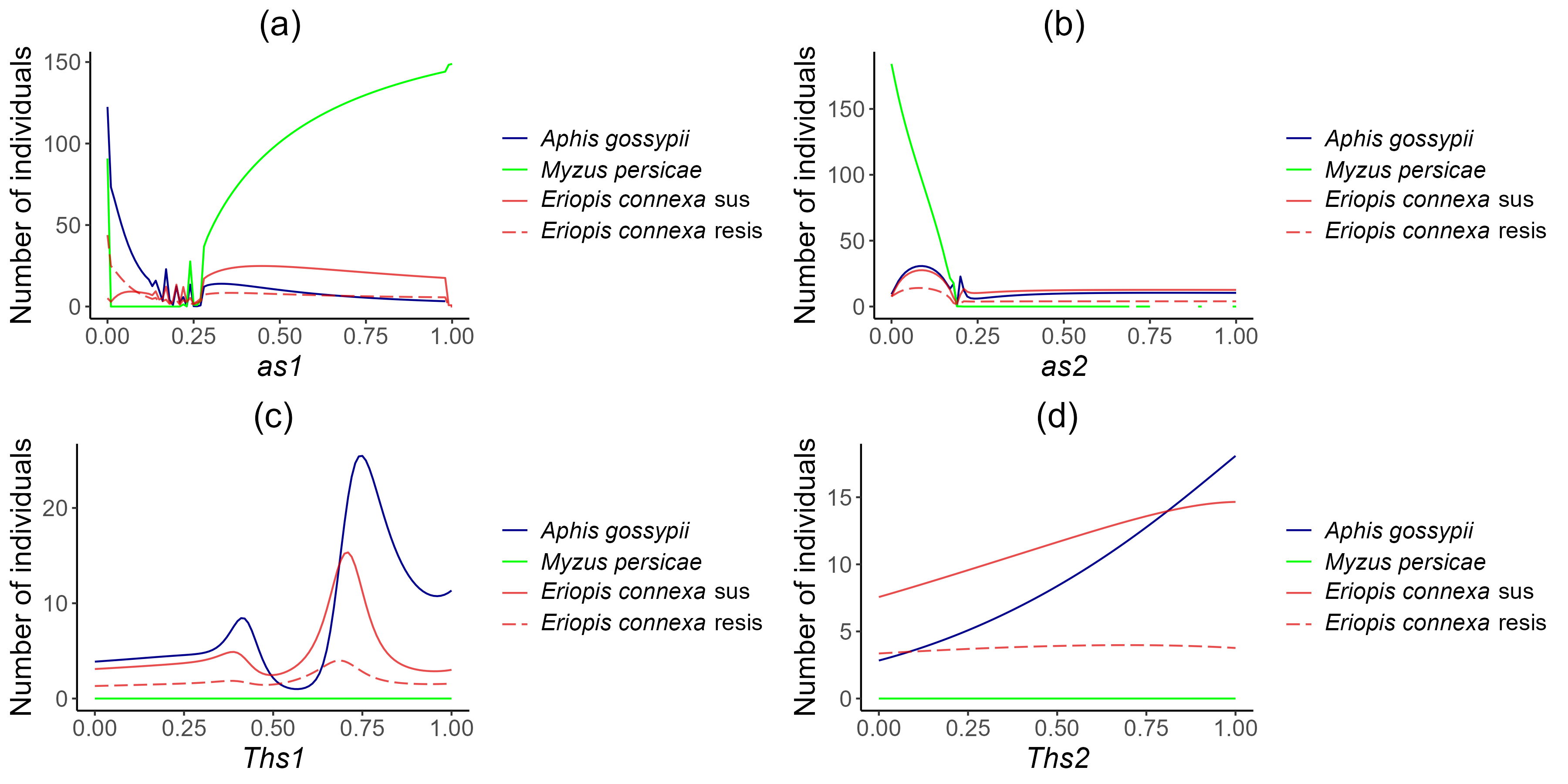} 
    \caption{Results of the changes for the predator parameters of the susceptible ladybird. Here, we selected the point ($N1$, $N2$, $P1$, $P2$) after 2000 models iterations for each parameter value of $a^{(s)}_1$, $a^{(s)}_2$, $T_{h_1}^{(s)}$, $T_{h_2}^{(s)}$. Initial conditions and parameter values: $N_1 = N_2 = 30$; $P_1 = 10$; $P_2 = 5$; $K_1 = K_2 = 200$;  $a^{(r)}_1 = 0.10$; $a^{(r)}_2 = 0.11$; $T_{h_1}^{(r)} = 0.62$; $T_{h_2}^{(r)} = 0.67$; $d_2 = d_3 = 0.05$; $g^{(s)}_1 = g^{(s)}_2 = 0.2$; $g^{(r)}_1 = g^{(r)}_2 = 0.17$.}
    \label{fig:image11}
\end{figure*}

The analysis of population dynamics, varying the attack rate ($as$) and handling time ($Ths$) parameters for susceptible and resistant \textit{E. connexa}, reveals distinct patterns in the behaviour of both prey populations (\textit{A. gossypii} and \textit{M. persicae}) and the predator across the simulations. \autoref{fig:image11} shows the dynamic conforming changes in the susceptible ladybird parameters (attack rate and handling time). When the range of attack rate ($a^{(s)}_1$) is between $0 < a^{(s)}_1 < 0.25$, a drastic decline in the density of \textit{M. persicae} (green line) is observed, while \textit{A. gossypii} (blue line) exhibits a gradual decline, followed by oscillations without a clear growth pattern. This suggests that the decline in \textit{M. persicae} allowed a small increase in \textit{A. gossypii}. However, after $a^{(s)}_1 \approx 0.25$,  the density of \textit{A. gossypii} entered a continuous decline and stayed with values close to zero. In contrast, the density of \textit{M. persicae} begins to increase significantly, suggesting that intense predation on \textit{A. gossypii} creates a competitive release for \textit{M. persicae}, allowing its uncontrolled growth. The population of susceptible \textit{E. connexa} (solid red line) remains stable at moderate levels, while the population of resistant \textit{E. connexa} (dashed red line) remains at even lower values (\autoref{fig:image11}a).

In the variation of $a^{(s)}_2$ (attack rate directed at \textit{M. persicae}), a significant reduction in the density of \textit{M. persicae} is observed as the parameter increases. Unlike the behaviour observed in $a^{(s)}_1$, the densities of \textit{A. gossypii}, and susceptible and resistant ladybirds remain relatively stable throughout the variation (\autoref{fig:image11}b). At low handling time ($T_{h_1}^{(s)}$) values, both \textit{A. gossypii} and predators (susceptible and resistant) had low density, while \textit{M. persicae} remained close to zero. As handling time ($T_{h_1}^{(s)}$) increases, an oscillatory behaviour is observed in the density of \textit{A. gossypii} and the predators. The aphid \textit{A. gossypii} has more pronounced peaks, while the susceptible and resistant \textit{E. connexa} populations show moderate growth trends. This oscillatory pattern suggests that increased handling time reduces predation efficiency, allowing \textit{A. gossypii} populations to recover periodically before being controlled again by \textit{E. connexa} (\autoref{fig:image11}c).

In the variation of handling time for \textit{M. persicae} ($T_{h_2}^{(s)}$), a continuous growth in the density of \textit{A. gossypii} is observed. The density of \textit{M. persicae} remains low throughout the variation of $T_{h_2}^{(s)}$. This suggests that the indirect competition between prey may be modulated by handling time where increased $T_{h_2}^{(s)}$ reduces predation on \textit{M. persicae}, favouring the population growth of \textit{A. gossypii}. The populations of susceptible and resistant \textit{E. connexa} demonstrate a stable behaviour, with a slight increase over time (\autoref{fig:image11}d).
 
\FloatBarrier

\begin{figure*} [ht]
    \centering
    \includegraphics[width = 0.8\textwidth]{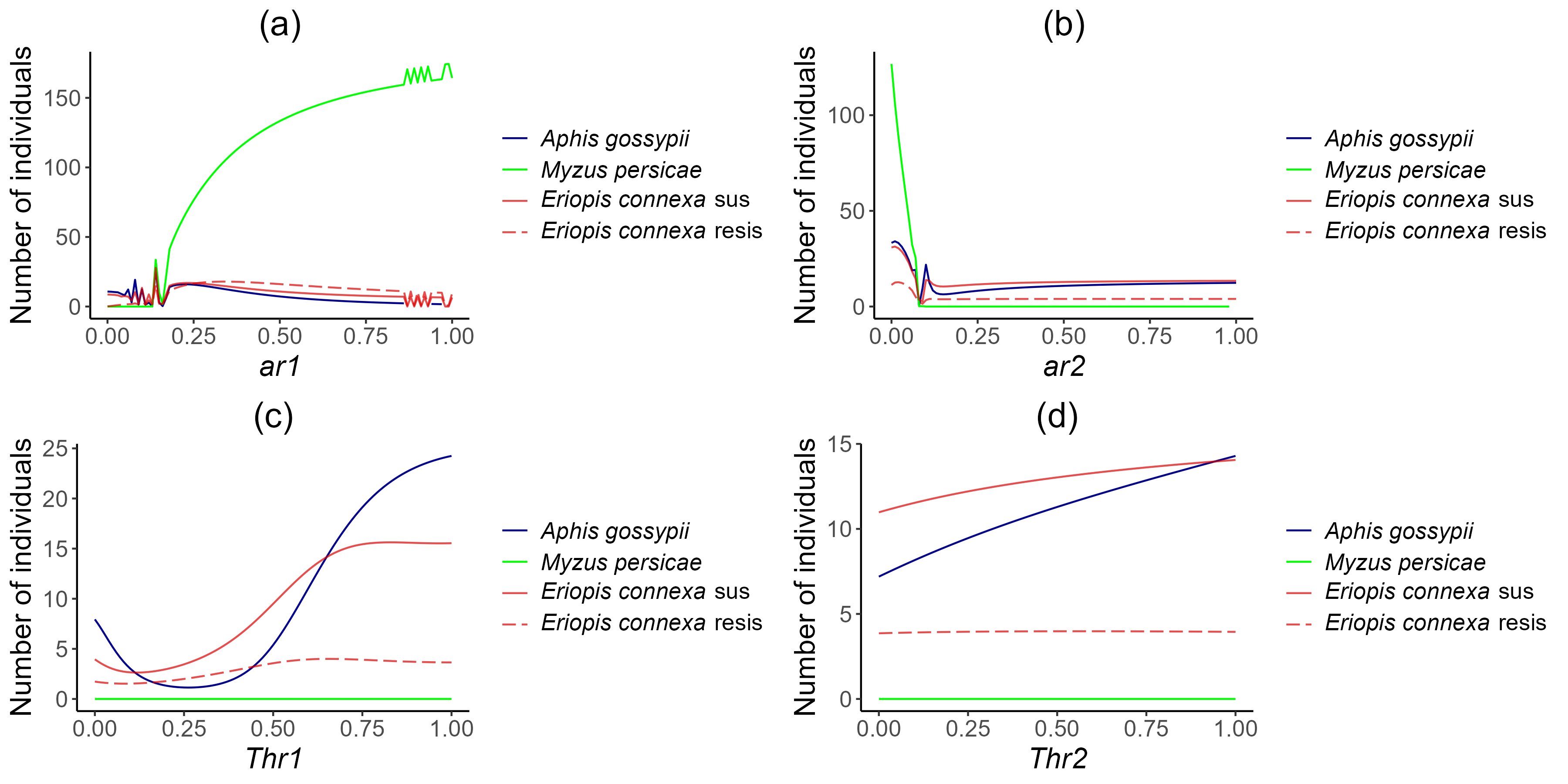} 
    \caption{Results of the changes for the predator parameters of the resistant ladybird.  Here, we selected the point ($N1$, $N2$, $P1$, $P2$) after 2000 models iterations for each parameter value of $a^{(r)}_1$, $a^{(r)}_2$, $T_{h_1}^{(r)}$, $T_{h_2}^{(r)}$. Initial conditions and parameter values: $N_1 = N_2 = 30$; $P_1 = 10$; $P_2 = 5$; $K_1 = K_2 = 200$; $a^{(s)}_1 = 0.20$; $a^{(s)}_2 = 0.21$; $T_{h_1}^{(s)} = 0.68$; $T_{h_2}^{(s)} = 0.73$; $d_2 = d_3 = 0.05$; $g^{(s)}_1 = g^{(s)}_2 = 0.2$; $g^{(r)}_1 = g^{(r)}_2 = 0.17$.}
    \label{fig:image12}
\end{figure*}

\autoref{fig:image12} shows the dynamic conforming changes in the resistant ladybird parameters. The attack rate ($a^{(r)}_1$ and $a^{(r)}_2$) and handling time ($T_{h_1}^{(r)}$ and $T_{h_2}^{(r)}$) values for resistant \textit{E. connexa} have distinct impacts on the population dynamics of prey (\textit{A. gossypii} and \textit{M. persicae}) and susceptible and resistant predators. For $a^{(r)}_1$ values between $0$ and $0.25$, it is observed that the density of \textit{M. persicae}(green) drops to zero quickly, suggesting that, even with low predatory efficiency on \textit{A. gossypii}, the dynamics of the system lead \textit{M. persicae} to extinction. This drop may be associated with indirect competition (interference), where the population of \textit{A. gossypii} remains at low levels, preventing \textit{M. persicae} from sustaining itself in the system. \textit{A. gossypii} gradually decreases with oscillations. 

After $a^{(r)}_1 \approx 0.18$, the increase in the attack rate leads to a continuous reduction in the density of \textit{A. gossypii}, while \textit{M. persicae} shows a significant increase due to competitive release since predation is focused on \textit{A. gossypii}. In this scenario, the density of predators \textit{E. connexa} susceptible and resistant shows a decrease. When $a^{(r)}_1$ is between $0.8$ and $0.9$ oscillations can be observed in \textit{M. persicae} and susceptible and resistant predators, indicating instability (\autoref{fig:image12}a). Initially, with low values of $a^{(r)}_2$ ($< 0.1$), populations of \textit{M. persicae} remain at high levels, indicating inefficient predation. Densities of \textit{A. gossypii} remains stable, and the ladybird has very low densities. The increase in attack rate ($> 0.1$) results in a drastic reduction in the density of \textit{M. persicae}, while \textit{A. gossypii} remains unchanged. The resistant \textit{E. connexa} population shows a slight recovery but is still at lower levels compared to susceptible predators (\autoref{fig:image12}b). 

With reduced handling times for \textit{A. gossypii} ($T_{h_1}^{(s)}$), the density of this species gradually decreases due to high predatory efficiency. In contrast, the density of the resistant ladybird increases moderately. The density of \textit{M. persicae} remains close to zero. Increasing handling times leads to a decrease in predatory efficiency, allowing \textit{A. gossypii} to increase significantly and after 0.75, starts to decrease again. At the same time, the density of susceptible and resistant \textit{E. connexa} increase and decrease following \textit{A. gossypii }(\autoref{fig:image12}c). When the value of handling time for \textit{M. persicae} ($T_{h_2}^{(s)}$) is low, there is high predatory efficiency, maintaining this species at extremely low densities. In contrast, \textit{A. gossypii} shows constant growth due to lower handling time values for \textit{M. persicae} than for \textit{A. gossypii}. When handling time increased, \textit{M. persicae} remained at very low densities while \textit{A. gossypii} grew continuously, benefiting from lower consumption by the resistant ladybird. Populations of resistant \textit{E. connexa} stabilize at low values, indicating lower predatory efficiency (\autoref{fig:image12}d). 

\FloatBarrier

\section{Discussion}
Our results demonstrate that parameters such as mortality, attack rate, and handling time are critical in shaping predator-prey interactions among \textit{E. connexa} and the aphids \textit{A. gossypii} and \textit{M. persicae}. We found that an increase in susceptible ladybird mortality ($d_2$) led to higher equilibrium populations of both aphid species ($N_1$ and $N_2$), suggesting that the reduced predation pressure allowed prey populations to grow and stabilize. In contrast, susceptible ladybirds ($P_1$) exhibited a decline as $d_2$ increased, while resistant ladybirds ($P_2$) showed a slight increase but remained at relatively low population levels. This indicates that while higher mortality of susceptible predators reduces their ability to control aphid populations, it does not necessarily create unfavourable conditions for resistant individuals. Instead, resistant ladybirds persist at low densities, likely benefiting from the reduced competition with susceptible individuals.

Bifurcation analyses revealed that increases in handling time reduce predatory efficiency while higher attack rates enhance the control of prey populations. Notably, we found that handling time also influences intraspecific competition among ladybirds, potentially favouring one population over another. When handling time increases, the predation efficiency of susceptible ladybirds declines, potentially giving an advantage to resistant individuals with shorter handling times. Conversely, when conditions are more favourable to susceptible ladybirds, the effect is reversed. 

The bifurcation analysis further demonstrated that stability transitions in the system occur in a non-linear manner, particularly for variations in the attack rate parameter. Instead of a simple shift between stable and unstable equilibria, the equilibrium solutions form a closed-loop structure, indicating cyclical transitions. This suggests the presence of hysteresis (where system behaviour depends on past states), where stability and instability depend not only on the current value of the attack rate but also on the system’s previous state. For instance, small changes in attack rate can result in the equilibrium moving between distinct regions. This implies that prey-predator dynamics under different attack rates may not shift gradually but rather exhibit abrupt transitions, potentially leading to ecological scenarios where populations undergo sudden shifts in dominance.

The study by \cite{lira2019predation}, which served as the basis for the predation rates of our model, reported that resistant ladybirds exhibited a roughly $46\%$ reduction in attack rate and an $8\%$ decrease in handling time compared to susceptible individuals. Based on our bifurcation analyses, we can hypothesize that shorter handling times may provide a competitive advantage to resistant ladybirds, allowing them to process prey more quickly. This advantage may become particularly significant under adverse conditions for susceptible ladybirds, such as scenarios involving insecticide applications, where resistant individuals will be selected and may outperform their counterparts because of their more efficient prey processing ability. 

Moreover, the existence of two equilibrium points—one stable and one unstable—within the same parameter range suggests that populations may experience alternate stable states depending on external disturbances or initial conditions. This reinforces the idea that the coexistence and competitive dynamics among ladybirds and aphids are not static but can vary under different ecological pressures. If handling time and attack rate thresholds govern these transitions, then even small environmental changes (such as insecticide exposure or fluctuations in prey availability) could shift population structures in unexpected ways.

In this way, our simulations with insecticide scenarios revealed significant ecological implications, where we observed that insecticide application affects aphid and ladybird populations differently. In scenarios with the absence of insecticide application (\autoref{fig:image10}a), susceptible ladybirds can control aphid populations, exhibiting typical predator-prey oscillations, while resistant ladybirds tend to become extinct because of adaptive costs associated with resistance \citep{ferreira2013life}. 

However, with the application of insecticides (\autoref{fig:image10}b and \autoref{fig:image10}c), the increase in predator mortality because insecticide generates a different scenario. After an initial decline in aphid populations, they quickly recover, reaching high densities. Also, pest resurgence after insecticide application has been widely documented in studies on aphids \citep{amad2003high, sial2018evaluation, ullah2020thiamethoxam, wang2023hormesis}. For instance, \cite{janssen2021pesticides} demonstrated through modelling that pest resurgence is expected even when effective natural enemies are present, even when these predators are less sensitive to insecticides than the pests. Pest resurgence can be caused by several factors, including an increased growth rate due to the hormesis effect, a phenomenon in which exposure to sublethal doses of stressors (e.g., insecticides, heavy metals) triggers a biphasic biological response: low doses stimulate physiological or reproductive processes, while higher doses suppress them \citep{calabrese2002defining}.  Concurrently, a reduction in the population of natural enemies can further amplify resurgence by disrupting top-down biocontrol \citep{guedes2016pesticide}. The reduction in natural enemies caused by insecticide exposure, allows prey populations to escape predator control, resulting in higher densities.

In scenarios with insecticide-resistant aphids (\autoref{fig:image10}c), the initial reduction in aphid populations is less pronounced, and their densities rapidly increase after applications, surpassing those observed in scenarios without insecticide application and with insecticide but without aphid resistance. In this context, aphid resistance contributes to this population increase, as mortality rates decrease with each application while the intrinsic growth rate of aphids increases. These findings align with existing literature, which highlights enhanced aphid growth rates and fecundity following insecticide application \citep{wang2022sulfoxaflor, kerns2000sublethal, sial2018evaluation, wang2018laboratory}. Furthermore, transgenerational effects were reported in studies by \cite{qiu2025implications} and \cite{zeng2016sublethal}, which found significant increases in fecundity and population growth parameters of \textit{A. gossypii} and \textit{M. persicae}, respectively, in the F1 generation after cyantraniliprole application.

For ladybirds, in the scenario without aphid resistance (\autoref{fig:image10}b), ladybird populations decline, especially the susceptible population because of the higher susceptibility of predators to the insecticide compared to aphids and resistant ladybird \citep{spindola2013survival, noelia2023toxicity, fogel2016toxicity}. Resistant populations gradually increase as susceptible populations decline, benefiting from their lower mortality rate under insecticide exposure. However, the simulated insecticide is not $100\%$ selective, leading to a $20\%$ increase in the mortality rate of resistant individuals. Despite this relatively low impact, resistant ladybirds are unable to reach high densities because of adaptive costs associated with the resistance phenotype. In \textit{E. connexa}, insecticide resistance is generally associated with fitness costs, including reduced fecundity \citep{rodrigues2020stability}, as well as impacts in predatory behaviours \citep{rodrigues2016ontogenic, lira2019predation, lira2023predation}, which reduce their predation efficiency.

Despite their low densities, resistant ladybirds still contribute to aphid population control. Furthermore, although there are peaks in aphid populations during applications, after these peaks, the density does not reach the same high levels as in untreated scenarios. \cite{spindola2013survival} found that resistant ladybirds have a survival rate of $82\%$ compared to $3\%$ in susceptible ones. The authors argue that even at low population densities, resistant \textit{E. connexa} that survive insecticide application are expected to help minimize aphid numbers, highlighting the importance of these findings for integrated pest management strategies.

In the scenario with aphid resistance (\autoref{fig:image10}c), the reduced mortality of aphids allows resistant ladybirds to maintain their presence over time. However, the aphid population remains high during the applications, but the substantial increase in resistant ladybirds observed during the final application causes a significative decline in the aphid population, reducing them to very low values. This suggests that, aphid populations initially grow due to reduced predation, but their population collapses as the resistant predator population expands. Furthermore, the presence of resistant ladybirds at higher densities enables them to remain in the system for longer periods, maintaining more controlled prey densities even after applications cease. These findings suggest that the presence of resistant predators contributes to pest control, especially under conditions where aphid resistance complicates management efforts.

Therefore, the model presented here suggests that the coexistence of susceptible and resistant ladybirds provides an "ecological insurance", where the presence of resistant individuals compensates for the high mortality rates observed in susceptible populations. These results indicate that introducing resistant predators can be an effective strategy for aphid control in environments where insecticide resistance has evolved. The delay in aphid decline suggests that there is a transitional period where resistant predators need time to establish themselves before effectively regulating the aphid population. This dynamic can be a useful strategy in integrated pest management, providing insights that can help in pest control. \cite{spindola2013survival} proposed the release of resistant \textit{E. connexa} in cotton crops based on their findings on insecticide selectivity. This approach appears viable, as the resistant phenotype is not entirely lost in the population in the absence of selection pressure, allowing for re-selection of resistance traits following insecticide applications \citep{nascimento2023heterosis}. Although our model does not account for ladybird releases, the simulations suggest that a gradual increase in resistant populations in the field could significantly enhance aphid control. This topic represents a promising area for future research, with the potential to optimize the use of resistant predators in pest management programs.

However, it is essential to consider potential trade-offs associated with resistance in predators. If resistant ladybirds exhibit reduced efficiency in prey consumption or reproductive capacity in the absence of insecticide pressure, their long-term persistence may be compromised. The resurgence of aphid populations in the final times highlights the long-term risk of pest recovery, which is probably because of the low recovery of susceptible ladybirds after insecticide applications, following the decrease in resistant ladybirds creating a competition scenario. Studies such as \cite{trumper1998modelling} have shown that the use of insecticides to reduce pests can also reduce predator populations, even in the absence of direct mortality. Therefore, while the presence of resistant ladybirds is relevant for pest control, it cannot resolve the issue by itself. The reduction in aphid populations also affects ladybirds, particularly because of adaptive costs associated with resistance and the reduction of the aphid population. This highlights the need for a balanced approach that considers both pest and predator dynamics to maintain effective long-term pest management.

\cite{janssen2021pesticides}, in their modelling of insecticide application scenarios with and without predators, emphasized that insecticides negatively impact pest densities when natural enemies are absent, but this effect is mitigated in the presence of predators. In our model, without the presence of resistant ladybirds, a similar pattern may be found because of the significant mortality of susceptible ladybirds. However, the presence of resistant ladybirds prevents the aphid population from growing excessively during and after insecticide applications. The study by \cite{nascimento2023heterosis} found that, in the absence of selection pressure, the resistance phenotype in \textit{E. connexa} was reduced but not completely lost. This suggests that maintaining a population of resistant ladybirds for a period after applications, until susceptible ladybirds recover can help to control the pest.

Another relevant consideration is that applying insecticides targeting non-aphid pests may indirectly influence aphid population dynamics. For example, the boll weevil, \textit{Anthonomus grandis} Boheman (Coleoptera: Curculionidae), is one of the most severe cotton pests, particularly in Brazil. Controlling boll weevils and other cotton pests, such as defoliators, still relies heavily on insecticide applications, particularly pyrethroids \citep{razaq2019insect, showler2007subtropical}. However, certain pyrethroids, such as lambda-cyhalothrin (LCT), exhibit ineffective efficacy against aphids, possibly leading to their proliferation or outbreaks \citep{dong2022heat, zambrano2021side, ohara2022profile}. Therefore,  maintaining populations of \textit{E. connexa} in cotton fields treated with LCT could be a valuable strategy for aphid control. The authors \cite{spindola2013survival} found that resistant \textit{E. connexa} populations showed promising potential for survival and persistence under LCT applications, reinforcing their suitability for integrated pest management (IPM) programs targeting boll weevil or other non-target pests of ladybird in cotton fields.

The population analysis at the end of the simulations, varying the attack rate and handling time parameters, showed again how these changes impact the system balance. Furthermore, in almost all scenarios, \textit{M. persicae} remained close to zero due to its lower growth rate compared to \textit{A. gossypii}. However, the results indicate that scenarios favouring its growth are possible.

Indirect interactions between prey mediated by the same predator population can result in apparent mutualism (positive effects) or apparent competition (negative effects) \citep{van2012prey, messelink2008biological}. Our results suggest that increasing the attack rate on \textit{A. gossypii} reduces the predation pressure on \textit{ M. persicae}, favouring its persistence and growth, which resembles an apparent mutualism. In contrast, increasing the attack rate on \textit{M. persicae} does not relieve \textit{A. gossypii} from predation, since ladybirds maintain a significant attack rate on \textit{A. gossypii}, preventing its population growth. As the attack rate on \textit{M. persicae}, its population is also consumed more intensively. Consequently, both aphid species are subject to increased predation, ultimately leading to the extinction of both populations, a scenario characterized by apparent competition.

\cite{costa2020occurrence}, in their modelling of scenarios involving two prey species and a single predator, found that, depending on the chosen parameter values, interactions could range from apparent mutualism to apparent competition. Their model suggests that increasing the carrying capacity of each prey species could initially lead to negative interactions (apparent competition), followed by positive interactions (apparent mutualism). Our results expand on this perspective by highlighting that the attack rate also plays a crucial role. Depending on the attack rate for one species, the effects on the other may be positive or negative.

Lastly, the results of this study highlight the importance of mathematical models as tools to understand predator-prey dynamics under different conditions. Although the data were obtained from literature studies, experimental validation of the model predictions can help to adjust the parameters and increase their practical applicability. Furthermore, future research could incorporate additional variables, such as climatic factors, spatial dynamics, or the release of ladybirds, as previously mentioned. This would expand the range of simulated scenarios and further deepen our understanding of ecological interactions.

For these reasons, this study enhances our understanding of the predation dynamics of \textit{E. connexa}, highlighting how variations in behavioural rates can impact prey control. The simulations emphasize the significance of sustainable agricultural practices, including reduced insecticide applications and product rotations, and showcase the potential of mathematical models to inform integrated pest management strategies in agricultural systems.

\section*{CRediT authorship contribution statement}

\textbf{Anna Mara Ferreira Maciel:} Conceptualization, Investigation, Funding acquisition, Methodology, Software, Writing – original draft, Writing – review \& editing.
\textbf{Gabriel Rodrigues Palma:} Methodology, Software, Writing – review \& editing
\textbf{Lucas dos Anjos:} Methodology, Software, Writing – review \& editing
\textbf{Lucas Santos Canuto:} Conceptualization, Writing – review \& editing
\textbf{Wesley Augusto Conde Godoy:} Conceptualization,  Methodology, Software, Supervision, Writing – review \& editing
\textbf{Rafael de Andrade Moral:} Conceptualization,  Methodology, Software, Supervision, Writing – review \& editing

\section*{Funding sources} 
This work was supported by the São Paulo Research Foundation (FAPESP) (Grant No. 2022/12222-7), as part of the São Paulo Advanced Research Center for Biological Control (SPARCBio), hosted at the Luiz de Queiroz College of Agriculture (ESALQ), University of São Paulo (USP), and sponsored by FAPESP, Koppert, and USP. The author also received a fellowship from the Brazilian Federal Foundation for Support and Evaluation of Graduate Education (CAPES), under the CAPES-PRINT program (Grant No. 88887.886799/2023-00). Also, this publication has resulted from research conducted with the financial support of Taighde Éireann – Research Ireland under Grant 18/CRT/6049.

\section*{Declaration of competing interest} 
The authors declare that they have no known competing financial interests or personal relationships
that could have appeared to influence the work reported in this paper.

\section*{Acknowledgements}
The authors acknowledge the helpful comments and suggestions made by the reviewers on an earlier version of this work. They also gratefully acknowledge the valuable feedback received during the research exchange in Ireland, especially from Professor Caroline Brophy's research group at Trinity College Dublin. The authors thank the Brazilian Federal Foundation for Support and Evaluation of Graduate Education (CAPES) for the scholarship support and the CAPES-PRINT program (Grant No. 88887.886799/2023-00) for making the international exchange possible.

\appendix
\section*{Appendix} \label{app:appendix}
\addcontentsline{toc}{section}{Appendix}
\renewcommand{\thefigure}{A.\arabic{figure}} 
\setcounter{figure}{0}

\subsection*{Bifurcation Analysis – Methodology}

The bifurcation analysis was conducted by varying the parameter $d_2$, representing the density-dependent mortality of the susceptible ladybird. This analysis was conducted to explore the system's sensitivity to this parameter. The decision to focus on $d_2$ was based on a few considerations. Firstly, there are no well-defined values for the density-dependent mortality parameters ($d_2$ and $d_3$) in the literature, requiring an exploratory approach. With that in mind, the choice to perform the bifurcation on $d_2$ was grounded in the fact that this parameter is directly related to the population dynamics of the susceptible ladybird, which plays a more active role in the chosen system because we selected a scenario with no insecticide disturbances. In this scenario, the resistant ladybird population exhibits limited growth, being present in the system primarily with adaptive costs. Therefore, keeping $d_3$ constant simplifies the analysis and allows a focused examination of $d_2$ effects. 

By varying $d_2$, it is possible to identify how changes in the density-dependent mortality of the susceptible ladybird affect population persistence and extinction, providing insights into the system's stability boundaries. By identifying critical thresholds or points of bifurcation, we were able to pinpoint transitions among different dynamic regimes, such as stable equilibrium, oscillatory behaviour, or a population collapse. 

Bifurcation analysis was also performed on the ladybirds' predation parameters, i.e., attack rate and handling time. Since the study focuses on ladybirds, both the parameters of the susceptible and resistant individuals were examined. This allowed for a comprehensive evaluation of how predation dynamics influence the system's behaviour.

To perform these analyses, numerical bifurcation of the model as a function of $d_2$ and the predation parameters ($a^{(s)}_1$, $a^{(s)}_2$, $a^{(r)}_1$, $a^{(r)}_2$, $T_{h_1}^{(s)}$, $T_{h_2}^{(s)}$, $T_{h_1}^{(r)}$, and $T_{h_2}^{(r)}$) was employed using the software package \texttt{XXPAUT} \citep{xxpaut}. This software calculates the equilibrium points of the non-linear differential equations in the model as parameters change. The analysed interval ranged from $0$ to $10$. However, the bifurcation results show that the system's dynamic behaviour was restricted to smaller intervals for each parameter analysed, with the x-axis values in the results representing the range effectively explored by the model.

Notably, we choose to perform the bifurcation analysis in scenarios without the application of insecticides because, in the system with the application, it would not be possible to conclude whether the effect would be due to a change in the parameters or a disturbance.

\subsection*{Bifurcation Analysis – Results}
\begin{figure*} [ht]
    \centering
    \includegraphics[width = 0.8\textwidth]{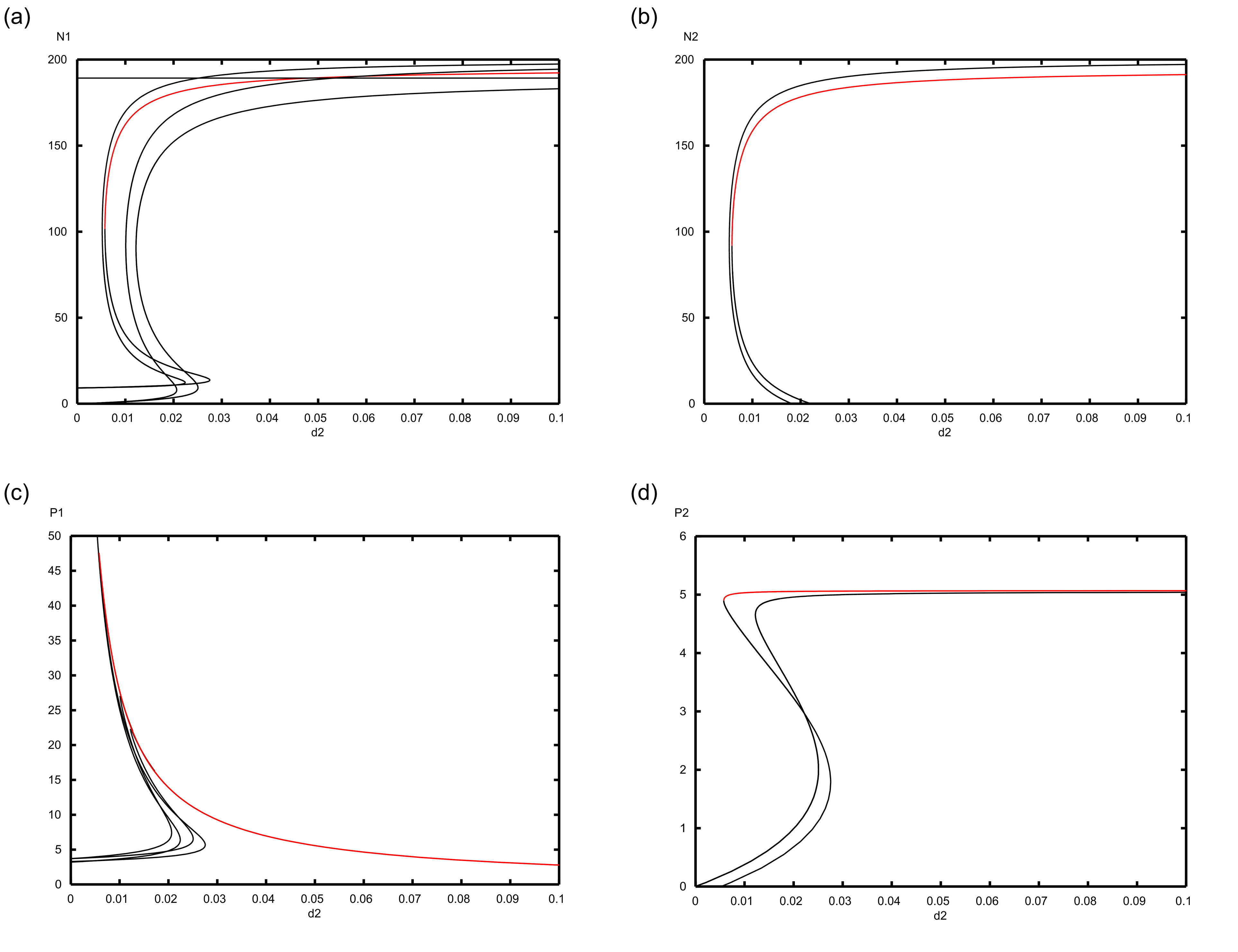} 
    \caption{Results of the bifurcation analysis of the population levels of prey and predator: (a) $N_1$, (b) $N_2$, (c) $P_1$, and (d) $P_2$ as a function of the parameter of susceptible ladybird mortality rate ($d_2$). The red lines indicate stable points of the system, while the black lines indicate points of instability. Initial conditions and parameter values: $N_1 102.2
$; $N2 = 92.5$; $P_1 = 47.6$; $P_2 = 4.9$; $K_1 = K_2 = 200$; $a^{(s)}_1 = 0.20$; $a^{(s)}_2 = 0.21$; $T_{h_1}^{(s)} = 0.68$; $T_{h_2}^{(s)} = 0.73$; $a^{(r)}_1 = 0.10$; $a^{(r)}_2 = 0.11$; $T_{h_1}^{(r)} = 0.62$; $T_{h_2}^{(r)} = 0.67$; $d_3 = 0.05$; $g^{(s)}_1 = g^{(s)}_2 = 0.2$; $g^{(r)}_1 = g^{(r)}_2 = 0.17$.}
    \label{fig:image1}
\end{figure*}

\autoref{fig:image1} represents the bifurcations generated for the mortality rate of the susceptible ladybird ($d_2$). 

\FloatBarrier

The first image (\autoref{fig:image1}a) represents the bifurcation for the species \textit{A. gossypii} ($N_1$) and  \textit{M. persicae} ($N_2$) (\autoref{fig:image1}a and \autoref{fig:image1}b). As $d_2$ increased, the equilibrium populations of $N_1$ and $N_2$ increased as well, suggesting that the increase in susceptible ladybird mortality ($d_2$) reduces the pressure that could limit these populations. Regarding susceptible ladybirds ($P_1$) (\autoref{fig:image1}c), as $d_2$ increased,  stability is observed, exhibiting a decrease in population. In contrast, resistant ladybirds ($P_2$) (\autoref{fig:image1}d) exhibited a slight increase, remaining at relatively low values.  

In general, the prey populations ($N_1$ and $N_2$) tended to grow and stabilize, while the ladybird populations ($P_1$ and $P_2$) either declined or stabilized at low levels (\autoref{fig:image1}).

\begin{figure*} [ht]
    \centering
    \includegraphics[width = 0.8\textwidth]{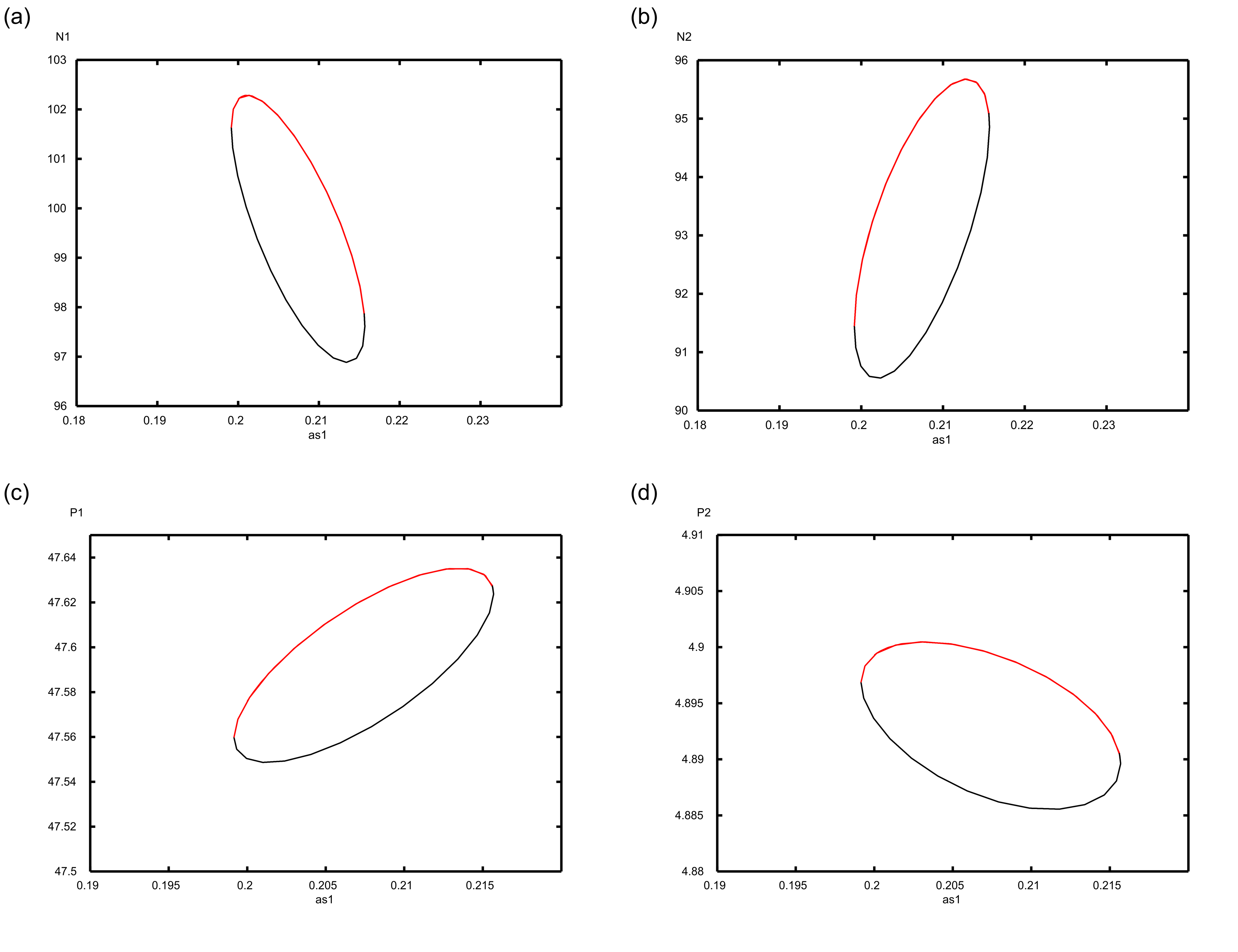} 
    \caption{Results of the bifurcation analysis of the population levels of prey and predator: (a) $N_1$, (b) $N_2$, (c) $P_1$, and (d) $P_2$ as a function of the parameter of susceptible ladybird attack rate on \textit{Aphis gossypii} ($a^{(s)}_1$). The red lines indicate stable points of the system, while the black lines indicate points of instability. Initial conditions and parameter values: $N_1 102.2
$; $N2 = 92.5$; $P_1 = 47.6$; $P_2 = 4.9$; $K_1 = K_2 = 200$; $a^{(s)}_2 = 0.21$; $T_{h_1}^{(s)} = 0.68$; $T_{h_2}^{(s)} = 0.73$; $a^{(r)}_1 = 0.10$; $a^{(r)}_2 = 0.11$; $T_{h_1}^{(r)} = 0.62$; $T_{h_2}^{(r)} = 0.67$; $d_3 = 0.05$; $d_2 = 0.0057$; $g^{(s)}_1 = g^{(s)}_2 = 0.2$; $g^{(r)}_1 = g^{(r)}_2 = 0.17$.}
    \label{fig:image2}
\end{figure*}

Regarding the bifurcation of ladybird predation parameters, \autoref{fig:image2} shows how the species respond to changes in the attack rate of the susceptible ladybirds ($P_1$) on \textit{A. gossypii} ($N_1$).

\FloatBarrier

The transition between stable (red) and unstable (black) equilibrium points occurs at specific parameter values. For each value of $a^{(s)}_1$, there is consistently one stable and one unstable equilibrium point, except at the transition points. As $a^{(s)}_1$ increases, the \textit{A. gossypii} ($N_1$) population stabilizes at intermediate attack rate values. However, the narrow stability interval shows that the system has an attractor in the phase space, leading the populations to oscillate in a predictable manner for certain values of $a^{(s)}_1$ beyond which equilibrium becomes unstable (\autoref{fig:image2}a). Similarly, the \textit{M. persicae} ($N_2$) population exhibits behaviour analogous to that of \textit{A. gossypii}, with stability confined to a limited range of  $a^{(s)}_1$  (\autoref{fig:image2}b).  Both the susceptible predator ($P_1$) (\autoref{fig:image2}c) and the resistant predator ($P_2$) \autoref{fig:image2}d) are also sensitive to variations on $a^{(s)}_1$. 

Overall, prey populations maintain higher density levels than ladybird populations, and within the ladybird group, the susceptible predator reaches higher densities than the resistant predator. 

\begin{figure*} [ht]
    \centering
    \includegraphics[width = 0.8\textwidth]{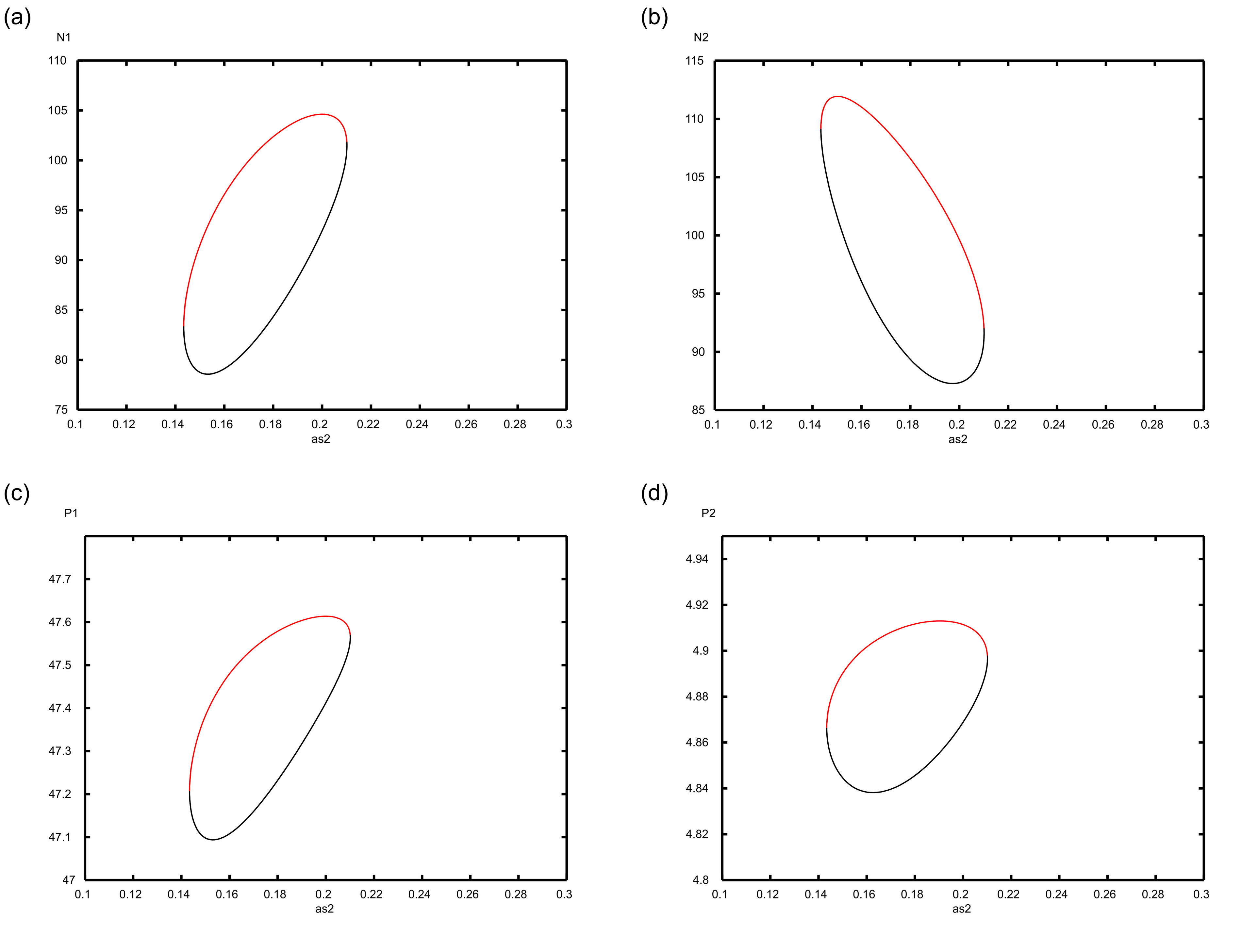} 
    \caption{Results of the bifurcation analysis of the population levels of prey and predator: (a) $N_1$, (b) $N_2$, (c) $P_1$, and (d) $P_2$ as a function of the parameter of susceptible ladybird attack rate on \textit{Myzus persicae} (${a^{(s)}_2}$). The red lines indicate stable points of the system, while the black lines indicate points of instability. Initial conditions and parameter values: $N_1 102.2
$; $N2 = 92.5$; $P_1 = 47.6$; $P_2 = 4.9$; $K_1 = K_2 = 200$;$a^{(s)}_1 = 0.20$; $T_{h_1}^{(s)} = 0.68$; $T_{h_2}^{(s)} = 0.73$; $a^{(r)}_1 = 0.10$; $a^{(r)}_2 = 0.11$; $T_{h_1}^{(r)} = 0.62$; $T_{h_2}^{(r)} = 0.67$; $d_3 = 0.05$; $d_2 = 0.0057$; $g^{(s)}_1 = g^{(s)}_2 = 0.2$; $g^{(r)}_1 = g^{(r)}_2 = 0.17$.}
    \label{fig:image3}
\end{figure*}

Regarding the dynamics for the attack rate parameter on \textit{M. persicae} ($a^{(s)}_2$) (A \autoref{fig:image3}), for \textit{A. gossypii} ($N_1$) (\autoref{fig:image3}a) and \textit{M. persicae} ($N_2$) (\autoref{fig:image3}b), the population decreases as $a^{(s)}_2$ increases. The stability interval follows the same pattern for $a^{(s)}s_1$. Both the susceptible predator ($P_1$) (\autoref{fig:image3}c) and the resistant predator ($P_2$) (\autoref{fig:image3}d) are also sensitive to variations on $a^{(s)}_2$. 

\begin{figure*} [ht]
    \centering
    \includegraphics[width = 0.8\textwidth]{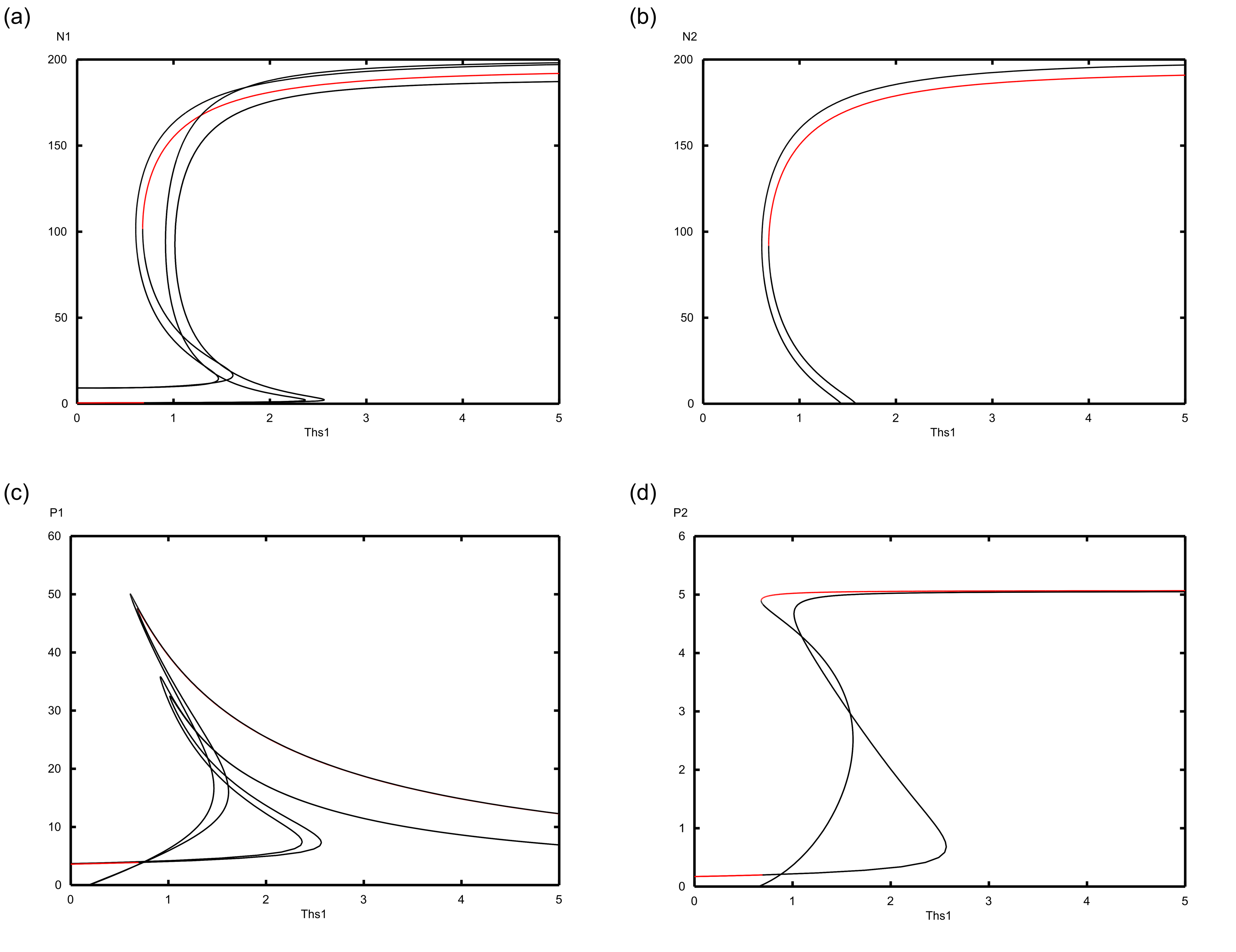} 
    \caption{Results of the bifurcation analysis of the population levels of prey and predator: (a) $N_1$, (b) $N_2$, (c) $P_1$, and (d) $P_2$ as a function of the parameter of susceptible ladybird handling time on \textit{Aphis gossypii} ($T_{h_1}^{(s)}$). The red lines indicate stable points of the system, while the black lines indicate points of instability. Initial conditions and parameter values: $N_1 102.2
$; $N2 = 92.5$; $P_1 = 47.6$; $P_2 = 4.9$; $K_1 = K_2 = 200$;$a^{(s)}_1 = 0.20$; $a^{(s)}_2 = 0.21$; $T_{h_2}^{(s)} = 0.73$; $a^{(r)}_1 = 0.10$; $a^{(r)}_2 = 0.11$; $T_{h_1}^{(r)} = 0.62$; $T_{h_2}^{(r)} = 0.67$; $d_3 = 0.05$; $d_2 = 0.0057$; $g^{(s)}_1 = g^{(s)}_2 = 0.2$; $g^{(r)}_1 = g^{(r)}_2 = 0.17$. }
    \label{fig:image4}
\end{figure*}

\FloatBarrier
The \autoref{fig:image4} shows the bifurcation for the handling time parameter of the susceptible ladybird ($T_{h_1}^{(s)}$) on \textit{A. gossypii}.

\FloatBarrier

As $T_{h_1}^{(s)}$ increases, the prey populations of \textit{A. gossypii} ($N_1$) (\autoref{fig:image4}a) and \textit{M. persicae} ($N_2$) (\autoref{fig:image4}b) also increase. The same behaviour occurs in resistant ladybird ($P_2$) (\autoref{fig:image4}d) while the susceptible ladybird ($P_1$) (\autoref{fig:image4}c) declines. 

We also can observe that  \autoref{fig:image4}a, \autoref{fig:image4}b, \autoref{fig:image4}c, and \autoref{fig:image4}d have two regions in red, exhibiting abrupt transition regions. This behaviour suggests a bifurcation, where the system abruptly shifts from one equilibrium state to another, resulting in a sudden change in population density — a phenomenon termed a branch. This indicates that, depending on the value of $T_{h_1}^{(s)}$, the population can exist in two distinct states. For example, in \autoref{fig:image4}d, it is observed that the $P_2$ population starts at low values and shows a slight increase as $T_{h_1}^{(s)}$ grows. However, upon reaching  $T_{h_1}^{(s)}$ $\approx$ $0.8$, an abrupt transition occurs, where $P_2$ jumps from approximately $0.2$ to $5$ in population density. 

In \autoref{fig:image4}c, two red regions are also present. Notably, the black and red lines are very close when the population of the susceptible ladybird ($P_1$) reaches $50$ individuals, but the red line takes precedence over the black one. This suggests that the populations will settle at the stable equilibrium point even though the unstable equilibrium is very close.

\begin{figure*} [ht]
    \centering
    \includegraphics[width = 0.8\textwidth]{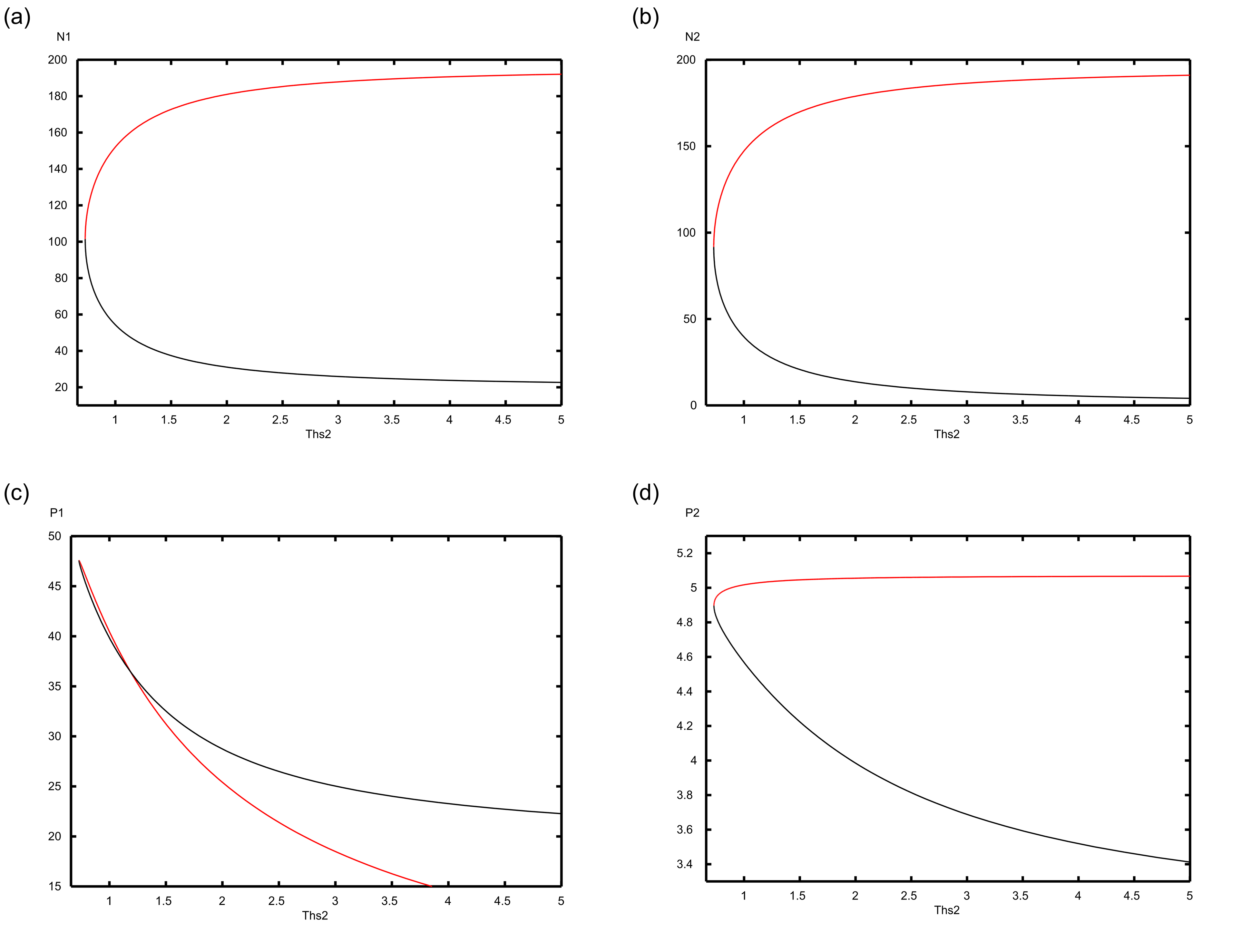} 
    \caption{Results of the bifurcation analysis of the population levels of prey and predator: (a) $N_1$, (b) $N_2$, (c) $P_1$, and (d) $P_2$ as a function of the parameter of susceptible ladybird handling time on \textit{Myzus persicae} (${T_{h_2}^{(s)}}$). The red lines indicate stable points of the system, while the black lines indicate points of instability. Initial conditions and parameter values: $N_1 102.2
$; $N2 = 92.5$; $P_1 = 47.6$; $P_2 = 4.9$; $K_1 = K_2 = 200$;$a^{(s)}_1 = 0.20$; $a^{(s)}_2 = 0.21$; $T_{h_1}^{(s)} = 0.68$; $a^{(r)}_1 = 0.10$; $a^{(r)}_2 = 0.11$; $T_{h_1}^{(r)} = 0.62$; $T_{h_2}^{(r)} = 0.67$; $d_3 = 0.05$; $d_2 = 0.0057$; $g^{(s)}_1 = g^{(s)}_2 = 0.2$; $g^{(r)}_1 = g^{(r)}_2 = 0.17$.}
    \label{fig:image5}
\end{figure*}

Regarding the handling time parameter of the susceptible ladybird for \textit{M. persicae} ($N_2$) (\autoref{fig:image5}), for \textit{A. gossypii} ($N_1$) (\autoref{fig:image5}a), \textit{M. persicae} ($N_2$) (\autoref{fig:image5}b),  and the resistant ladybird ($P_2$) (\autoref{fig:image5}d), as $T_{h_2}^{(s)}$ increases, two distinct equilibrium states emerge: a low population equilibrium (black line, unstable) and a high population equilibrium (red line, stable). A different pattern is observed in (\autoref{fig:image5}c) for susceptible ladybirds ($P_1$), where the population declines as $T_{h_2}^{(s)}$ increases. We also can see an inversion in the position of the stability line after close to $1.2$. The presence of both stable and unstable regions suggests that a threshold effect may govern the transition between states.

\begin{figure*} [ht]
    \centering
    \includegraphics[width = 0.8\textwidth]{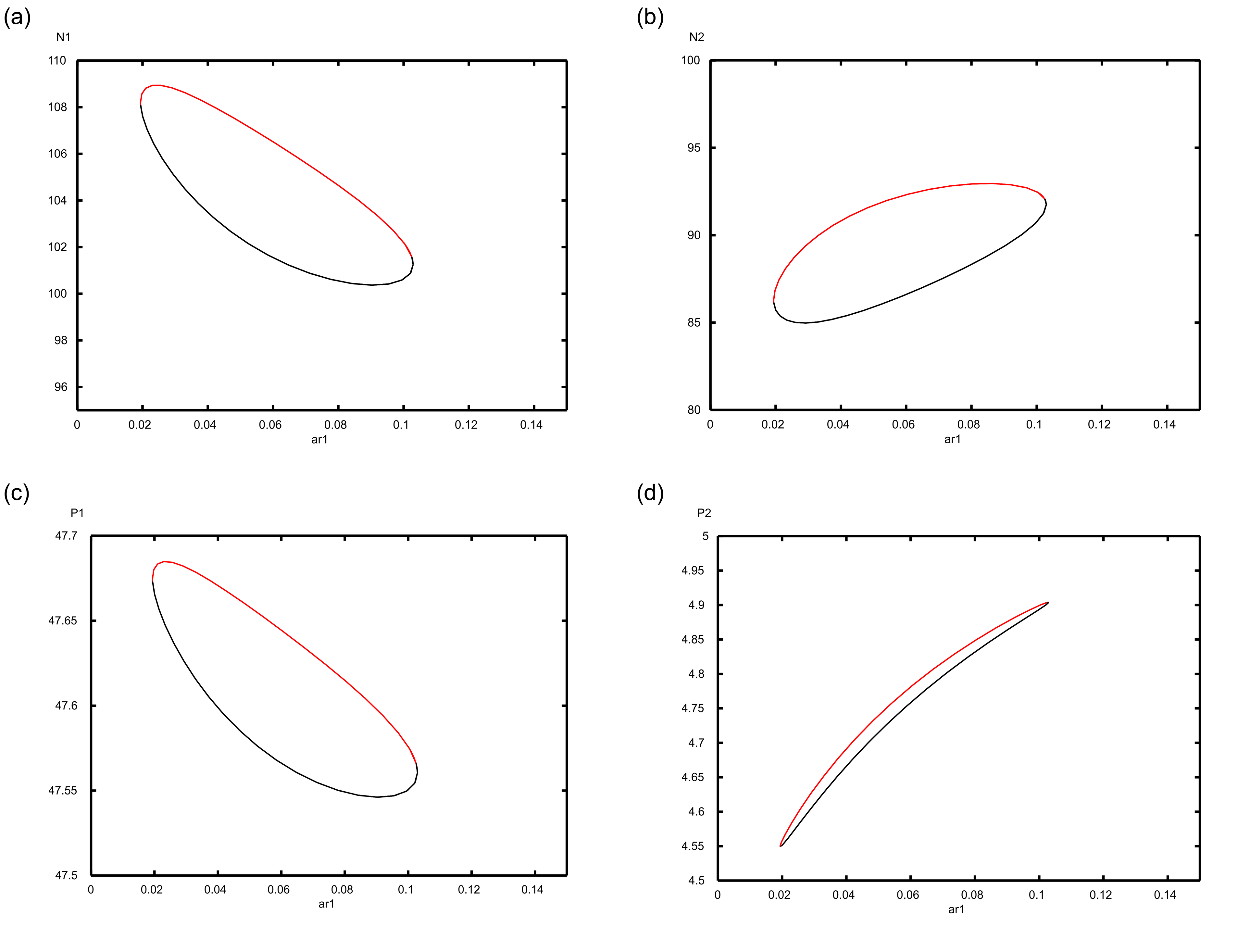} 
    \caption{Results of the bifurcation analysis of the population levels of prey and predator: (a) $N_1$, (b) $N_2$, (c) $P_1$, and (d) $P_2$ as a function of the parameter of resistant ladybird attack rate on \textit{Aphis gossypii} (${a^{(r)}_1}$). The red lines indicate stable points of the system, while the black lines indicate points of instability. Initial conditions and parameter values: $N_1 102.2
$; $N2 = 92.5$; $P_1 = 47.6$; $P_2 = 4.9$; $K_1 = K_2 = 200$;$a^{(s)}_1 = 0.20$; $a^{(s)}_2 = 0.21$; $T_{h_1}^{(s)} = 0.68$; $T_{h_2}^{(s)} = 0.73$; $a^{(r)}_2 = 0.11$; $T_{h_1}^{(r)} = 0.62$; $T_{h_2}^{(r)} = 0.67$; $d_3 = 0.05$; $d_2 = 0.0057$; $g^{(s)}_1 = g^{(s)}_2 = 0.2$; $g^{(r)}_1 = g^{(r)}_2 = 0.17$.}
    \label{fig:image6}
\end{figure*}

\FloatBarrier
For the system's behaviour in response to changes in the predation parameters of resistant ladybirds, for the attack rate parameter ($a^{(r)}_1$), \textit{A. gossypii} ($N_1$) (\autoref{fig:image6}a), \textit{M. persicae} ($N_1$) (\autoref{fig:image6}b), susceptible ladybird population ($P_1$) (\autoref{fig:image6}c), and the resistant ladybird population ($P_2$) (\autoref{fig:image6}d), the bifurcations show a closed curve similar with the attack rate parameter for susceptible ladybird ($a^{(s)}_1$ and $a^{(s)}_1$) (\autoref{fig:image2} and \autoref{fig:image3}).

\begin{figure*} [ht]
    \centering
    \includegraphics[width = 0.8\textwidth]{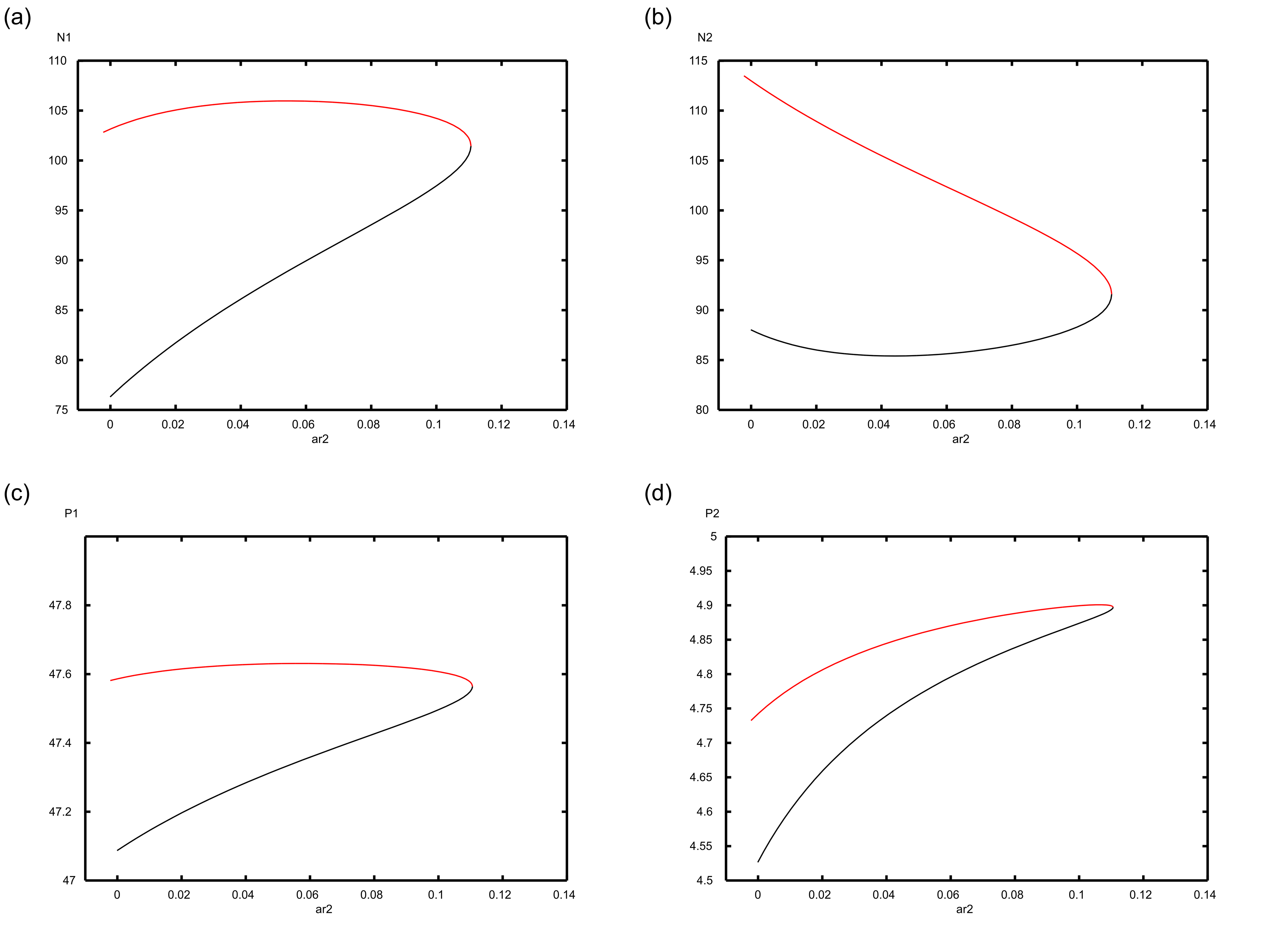} 
    \caption{Results of the bifurcation analysis of the population levels of prey and predator: (a) $N_1$, (b) $N_2$, (c) $P_1$, and (d) $P_2$ as a function of the parameter of resistant ladybird attack rate on \textit{Myzus persicae} ($a^{(r)}_2$). The red lines indicate stable points of the system, while the black lines indicate points of instability. Initial conditions and parameter values: $N_1 102.2
$; $N2 = 92.5$; $P_1 = 47.6$; $P_2 = 4.9$; $K_1 = K_2 = 200$;$a^{(s)}_1 = 0.20$; $a^{(s)}_2 = 0.21$; $T_{h_1}^{(s)} = 0.68$; $T_{h_2}^{(s)} = 0.73$; $a^{(r)}_1 = 0.10$; $T_{h_1}^{(r)} = 0.62$; $T_{h_2}^{(r)} = 0.67$; $d_3 = 0.05$; $d_2 = 0.0057$; $g^{(s)}_1 = g^{(s)}_2 = 0.2$; $g^{(r)}_1 = g^{(r)}_2 = 0.17$.}
    \label{fig:image7}
\end{figure*}

\FloatBarrier
The attack rate parameter of the resistant ladybird ($a^{(r)}_2$) on \textit{M. persicae} ($N_2$) (\autoref{fig:image7}) shows differences with two possible population states, as indicated by the attractor dynamics involving all species. 
 
\FloatBarrier

For \textit{A. gossypii} ($N_1$) (\autoref{fig:image7}a) population growth as $a^{(r)}_2$ increased. Stability occurs at high population levels, between approximately $100$ and $105$.  For \textit{M. persicae} ($N_2$), which is directly affected by the parameter, equilibrium pest density decreases in the interval approximately between $0$ and $0.1$ (\autoref{fig:image7}b). On the other hand, equilibrium pest density increases with the high values of $a^{(r)}_2$ in the interval $0.1$ and $0.12$ (\autoref{fig:image7}b).

For both susceptible ($P_1$) (\autoref{fig:image7}c) and resistant ($P_2$) (\autoref{fig:image7}d) ladybird populations, the equilibrium density increases as $a^{(r)}_2$ increases, but at high values ($> 0.1$), a transition occurs, starting to decrease. The population levels are small compared to $N_1$ and $N_2$, mainly for the resistant ladybird ($P_2$).

\begin{figure*} [ht]
    \centering
    \includegraphics[width = 0.8\textwidth]{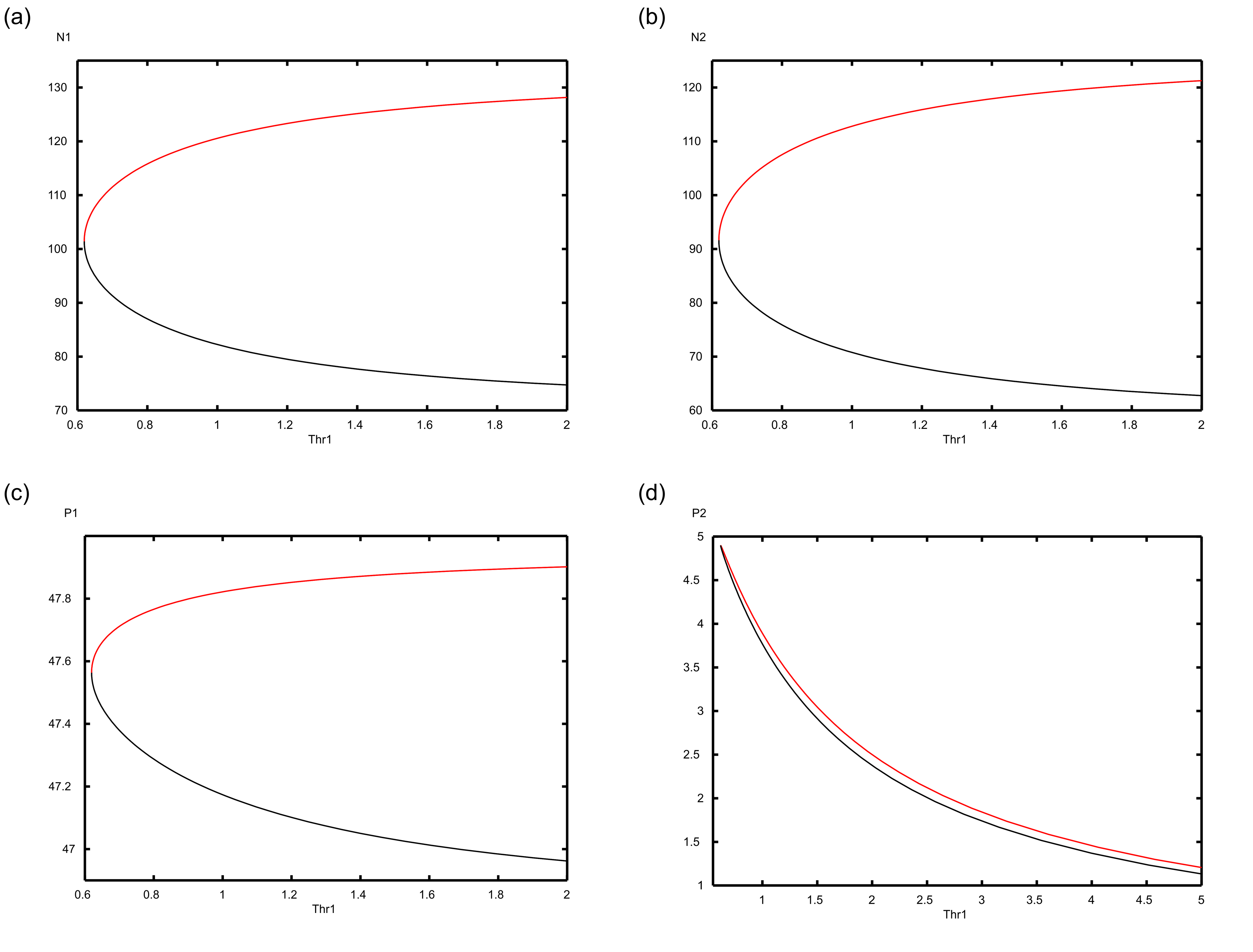} 
    \caption{Results of the bifurcation analysis of the population levels of prey and predator: (a) $N_1$, (b) $N_2$, (c) $P_1$, and (d) $P_2$ as a function of the parameter of resistant ladybird handling time on \textit{Aphis gossypii} (${T_{h_1}^{(r)}}$). The red lines indicate stable points of the system, while the black lines indicate points of instability. Initial conditions and parameter values: $N_1 102.2
$; $N2 = 92.5$; $P_1 = 47.6$; $P_2 = 4.9$; $K_1 = K_2 = 200$;$a^{(s)}_1 = 0.20$; $a^{(s)}_2 = 0.21$; $T_{h_1}^{(s)} = 0.68$; $T_{h_2}^{(s)} = 0.73$; $a^{(r)}_1 = 0.10$; $a^{(r)}_2 = 0.11$; $T_{h_2}^{(r)} = 0.67$; $d_3 = 0.05$; $d_2 = 0.0057$; $g^{(s)}_1 = g^{(s)}_2 = 0.2$; $g^{(r)}_1 = g^{(r)}_2 = 0.17$.}
    \label{fig:image8}
\end{figure*}

\begin{figure*} [ht]
    \centering
    \includegraphics[width = 0.8\textwidth]{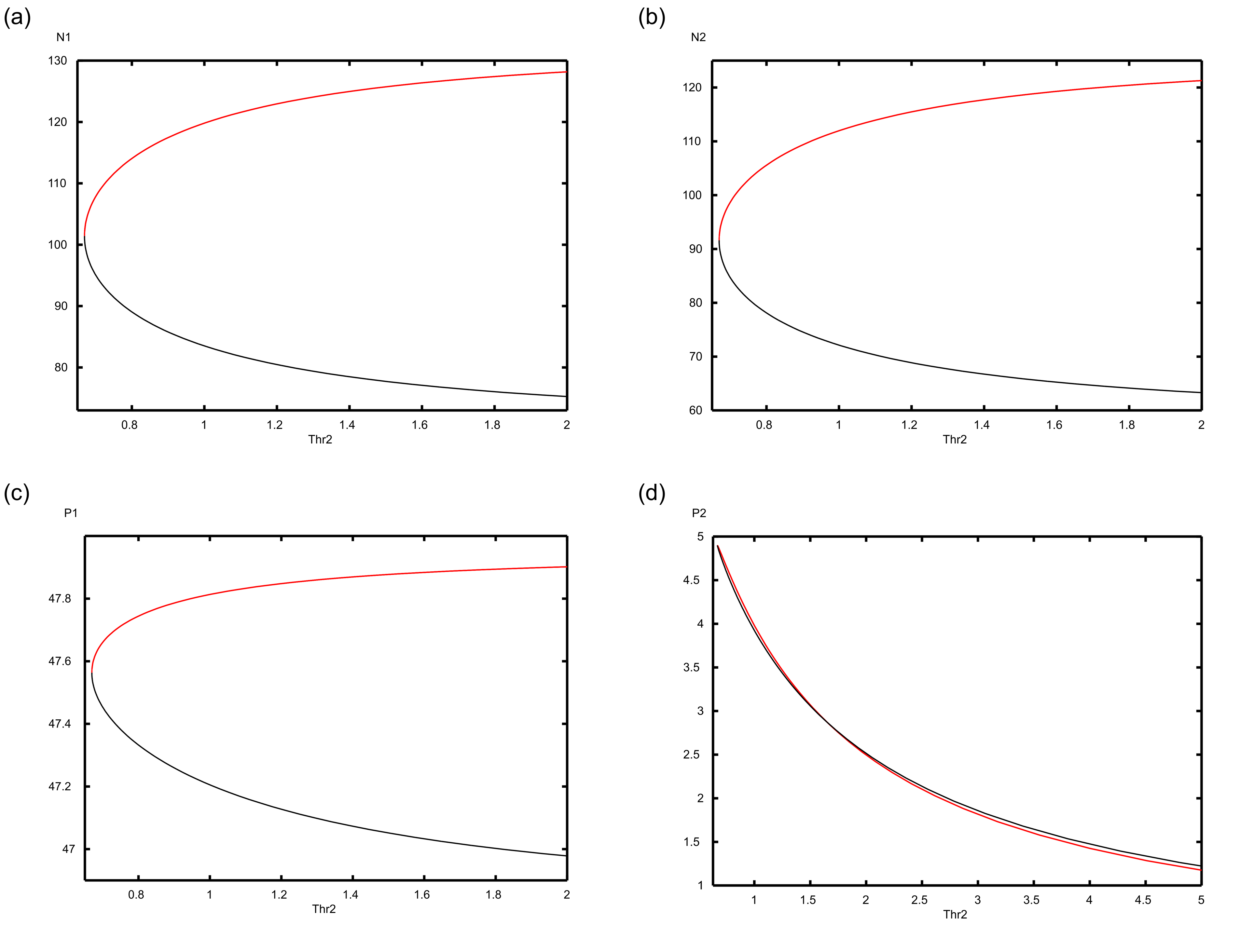} 
    \caption{Results of the bifurcation analysis of the population levels of prey and predator: (a) $N_1$, (b) $N_2$, (c) $P_1$, and (d) $P_2$ as a function of the parameter of resistant ladybird handling time on \textit{Myzus persicae} (${T_{h_2}^{(r)}}$). The red lines indicate stable points of the system, while the black lines indicate points of instability. Initial conditions and parameter values: $N_1 102.2
$; $N2 = 92.5$; $P_1 = 47.6$; $P_2 = 4.9$; $K_1 = K_2 = 200$;$a^{(s)}_1 = 0.20$; $a^{(s)}_2 = 0.21$; $T_{h_1}^{(s)} = 0.68$; $T_{h_2}^{(s)} = 0.73$; $a^{(r)}_1 = 0.10$; $a^{(r)}_2 = 0.11$; $T_{h_1}^{(r)} = 0.62$;$d_3 = 0.05$; $d_2 = 0.0057$; $g^{(s)}_1 = g^{(s)}_2 = 0.2$; $g^{(r)}_1 = g^{(r)}_2 = 0.17$.}
    \label{fig:image9}
\end{figure*}

For the handling time of resistant ladybirds ($T_{h_1}^{(r)}$) on \textit{A. gossypii} ($N_1$) and on \textit{M. persicae} $T_{h_2}^{(r)}$ (\autoref{fig:image8} and \autoref{fig:image9}), For aphids and susceptible ladybird as $T_{h_1}^{(r)}$ and $T_{h_2}^{(r)}$ increases, unstable equilibrium occurs at low levels of population and stable equilibrium emerges at high levels of population. 

In contrast, the resistant ladybird ($P_2$) decreases as $T_{h_1}^{(r)}$ and $T_{h_1}^{(r)}$ increases. Also, we can see that the equilibrium stability is very close to the instability, showing that this population is highly sensitive to this parameter (\autoref{fig:image8}d and \autoref{fig:image9}d).

\FloatBarrier

\FloatBarrier
\bibliographystyle{elsarticle-harv} 
\bibliography{referencias}

\end{document}